# Unveiling the Milky Way dust extinction curve in 3D

**Authors:** Xiangyu Zhang[1*†], Gregory M. Green[1†]

**Affiliations:**

[1]Galaxies and Cosmology, Max Planck Institute for Astronomy; Königstuhl 17, D-69120 Heidelberg, Germany

*Corresponding author. Email: xzhang@mpia.de

†These authors contributed equally to this work.

**Abstract:** Interstellar dust is a major foreground contaminant for many observations and a key component in the chemistry of the interstellar medium, yet its properties remain highly uncertain. Using low-resolution spectra, we accurately measure the extinction curve – a diagnostic of the grain properties – for 130 million stars, orders of magnitude more than previously available, allowing us to map its variation in the Milky Way and Magellanic Clouds in 3D in unprecedented detail. We find evidence that accretion is the dominant mechanism of grain growth in moderately dense regions, with coagulation dominating at higher densities. Moreover, we find that the extinction curve flattens in star-forming regions, possibly caused by cycling of large grains formed in molecular clouds, or by preferential destruction of small grains by supernova shocks.

**Main Text:**

Interstellar dust plays a major role in many areas of astronomy, either as a nuisance or as an object of study in its own right. Dust plays an important role in the chemistry of the interstellar medium (ISM), by – for example – exchanging metals with the gas phase, catalyzing the formation of $H_2$, and shielding dense molecular cloud cores from the interstellar radiation field. In the ultraviolet (UV), optical and near infrared (NIR), dust scatters and absorbs light, thus dimming, reddening and even polarizing background sources. Observations in many fields of astronomy must be corrected for these extinction effects. In the far infrared, dust emits thermally, creating a foreground contaminant for cosmic microwave background observations.

Given its centrality to so many areas of astronomy, the present relative lack of knowledge of interstellar dust composition is striking. Various observations have placed constraints on the composition and properties of the dust. To name a few: the UV extinction feature at $\sim 2175$ Å indicates the presence of a dust component with benzene-like carbon rings, such as graphite or polycyclic aromatic hydrocarbons; gas-phase elemental abundances in the ISM place constraints on the amount of each element that is locked up in the dust grains; the shape of the wavelength-vs.-extinction curve depends on both the optical properties of the dust grains and on their size distribution.

Until relatively recently, studies of the dust extinction curve have focused on small samples of sightlines, using high-resolution spectroscopy of O- and B-stars to measure extinction as a function of wavelength along hundreds of sightlines. These studies have revealed that in the Milky Way, the optical-NIR extinction curve along different sightlines can be approximately described by a single-parameter family of curves, typically parameterized by $R(V) \equiv A(V) \,/\, [A(B) - A(V)]$, which describes the slope of the extinction curve in optical wavelengths (*1*–*3*). With the advent of large spectroscopic and multiband photometric surveys, it has become possible to measure the extinction curve for a much larger number of sightlines. Schlafly et al. (*4*) combined spectroscopically determined stellar parameters with broadband photometry to determine the reddenings of $\sim 37{,}000$ stars in multiple colors, and then used these reddenings to estimate $R(V)$. Schlafly et al. (*5*) followed up on this work by mapping variation of dust $R(V)$ in three dimensions in the Milky Way, unexpectedly finding variations on kiloparsec scales. Zhang et al. (*6*, no relation) recently used a similar method to measure $R(V)$ for $\sim 3$ million stars, an increase of two orders of magnitude over the earlier catalog of Schlafly et al. (*4*).

In this work, we leverage low-resolution Gaia XP spectra (*7–10*) to obtain high-quality measurements of dust $R(V)$ for $\sim 130$ million stars, more than an order of magnitude more than have been available to date. The XP spectra are ideal for mapping $R(V)$ for a number of reasons. There are $\sim 220$ million sources with XP spectra – orders of magnitude more than in previous spectral catalogs, and approaching photometric surveys in sheer numbers of sources. In contrast to broadband photometry, XP spectra are able to precisely constrain stellar parameters (e.g., *11–13*). XP spectra are flux calibrated, allowing a direct measurement of extinction (*10*). Finally, high-quality *Gaia* parallax measurements are available for a large fraction of stars with XP spectra, which greatly aids in the disentangling parameters such as distance, extinction and surface gravity. We build on the forward model of XP spectra (augmented with near-infrared photometry) developed in Zhang et al. (*11*), which used a universal extinction curve. In this work, we allow the shape of the model extinction curve to vary, which enables us to measure dust $R(V)$ for individual stars. As in our previous work, we learn both our stellar model and the dust extinction curve directly from the data, using a small subset (approximately $1\%$) of the Gaia XP catalog that has matching higher-resolution LAMOST spectroscopy. We then apply this model to all $\sim 220$ million Gaia XP spectra to infer both stellar and dust properties.

Our work makes it possible to trace dust $R(V)$ throughout a significant volume of the Milky Way and in the Magellanic Clouds with unprecedented spatial resolution and coverage. Our results uncover a wealth of structure that models of dust composition and evolution will have to explain:

- We see intriguing hints that accretion is the dominant process of dust growth in moderate-density environments, causing $R(V)$ to decrease, while coagulation is the dominant process of grain growth in high-density environments, causing $R(V)$ to again increase.

- We see strong correlation between high-$R(V)$ regions and star-forming regions, as indicated by O/B-stars in the Galactic disk within 2.6 kpc, and by well known star-forming regions in the Large Magellanic Cloud (LMC).

## Stellar model and parameters

In order to infer the extinction of every *Gaia* XP star, we build a forward model that predicts flux as a function of stellar parameters: stellar type $(\theta \equiv T_{\rm eff},\ [{\rm Fe/H}],\ \log g)$, parallax ($\varpi$), a scalar amount of extinction ($E$), and a parameter that determines the slope of the extinction curve ($\xi$). The intrinsic stellar spectrum is described by a neural network, $\vec{F}_\lambda(\theta)$, while distance and extinction are treated in a physically informed manner. The structure of the neural network of this model is described in detail in Appendix D and Fig. D5. We implement the model in an auto-differentiable framework, TensorFlow 2 (*14*).

Our model can be summarized by the equations

$$\vec{f}_\lambda(\theta, \varpi, E, \xi) = \vec{F}_\lambda(\theta)\varpi^2 \exp\left[-E\vec{R}(\xi)\right], \quad [1]$$

$$\text{where } \ln \vec{R}(\xi) = \ln \vec{R}_0 + \tanh(\xi)\, \Delta \ln \vec{R}\,. \quad [2]$$

Our previously published model of XP spectra (*11*, hereafter, “v1”) made the simplifying assumption that there exists a universal extinction curve shared by all stars. In contrast, here we allow the slope of the extinction curve to vary from star to star within a one-parameter family (Eq. 2). We do not impose prior knowledge on the extinction curve ($\ln \vec{R}_0$ and $\Delta \ln \vec{R}$), but rather apply a simple regularization term that encourages continuity as a function of wavelength (See Appendix D.2 for details). The parameter $\xi$ controls the shape of the extinction curve, functioning similarly to $R(V)$ in other dust models (e.g., *1–3*). We apply a tanh function to the parameter $\xi$ in order to limit the range of extinction-curve variation.

We train this model using stellar-type parameter estimates from LAMOST DR8 (*15*, *16*) and the “Hot Payne” catalog (*17*) as priors on $\theta$, Bayestar19 (*18*) reddening estimates as priors on E, Gaia DR3 (*8*) parallax measurements as priors on $\varpi$, and a unit normal prior on $\xi$. For stars not covered by Bayestar19, we refer to the 2D dust map “SFD” (*19*). The XP spectral flux is represented in wavelength space using the same sampling as in v1 (*11*), together with near-infrared photometric bands J, H, and Ks from 2MASS (*20*), and W1 and W2 from unWISE (*21*). During training, we simultaneously update the stellar parameters and the model, using a combination of the priors and the likelihood of the observed flux. We describe our training procedure in detail in Appendix D.

## Results for $R(V)$

We obtain 130 million reliable $R(V)$ estimates across the entire sky, enabling us to explore the variation of $R(V)$ in 3D in the Milky Way galaxy in unprecedented detail.

We consider stars that simultaneously meet the following standards as having reliable extinction-curve measurements: $E > 0.1$, $\xi_{\mathrm{err}} < 0.2$ and good observational and fitting quality (See Appendix D.4 for details). With this standard, our catalog contains 130 million reliable measurements of $\xi$.

Across a wide range of $R(V)$ values, our learned family of extinction curves matches that of Gordon et al. (*3*, “G23”), which is based on higher-resolution spectroscopic data, to within $\sim 5\%$ (See Appendix A for more details). In addition, we compare our inferred $R(V)$ values for individual stars (as calculated from our extinction-curve parameter, $\xi$; See Appendix B and Eqs. B7 and B8 for details) with existing catalogs. Our per-star $R(V)$ results agree well with existing (but far smaller) catalogs based on high-resolution spectroscopy in the sense of having a monotonic correspondence. Exact $R(V)$ values are not necessarily identical because of different definitions of $R(V)$, but can be easily converted by a monotonic function. Our $R(V)$ values are high (low) where other catalogs indicate high (low) values. Since most of the stars in the Zhang et al. (*6*, no relation; hereafter “ZYC23”) have XP spectra, we manage to reproduce the 2D distribution from ZYC23. We show the comparison with Schlafly et al. (*4*, hereafter “S16”) and ZYC23 in Appendix E.

We apply a relatively simple binning method to our per-star extinction estimates to create a 3D map of $R(V)$. In particular, we calculate maps of average $A(V)$ and $A(B)$ across the sky on a grid of distances, and then calculate $R(V)$ within each distance bin using the difference in B and V -band extinction in consecutive distance slices: $R(V) \equiv \Delta A(V) / \Delta E(B - V)$. We use high-extinction ($E > 0.1$) stars that pass observational and fitting quality cuts (See Appendix D.4 for details). Despite the simplicity of this method, it nevertheless reveals rich information about the 3D variation in the dust extinction curve.

In Fig. 1, we show a bird's-eye view of $R(V)$ in the Galactic plane ($|z| < 400$ pc) within 2.6 kpc from the Sun ($\odot$). We find spatial correlations between $R(V)$ and several local large structures and arms of the Milky Way, such as the Radcliffe Wave (*26*), the Split (*27*) and Carina-Sagittarius. The Carina-Sagittarius arm shows higher-than-average $R(V)$, while most of the Split has lower $R(V)$, although they are both concentrated regions of molecular clouds. This may be due to their differences in environment. For example, Carina-Sagittarius hosts a relatively high density of O/B-stars and star-forming regions, which are lacking in the Split (*27*).

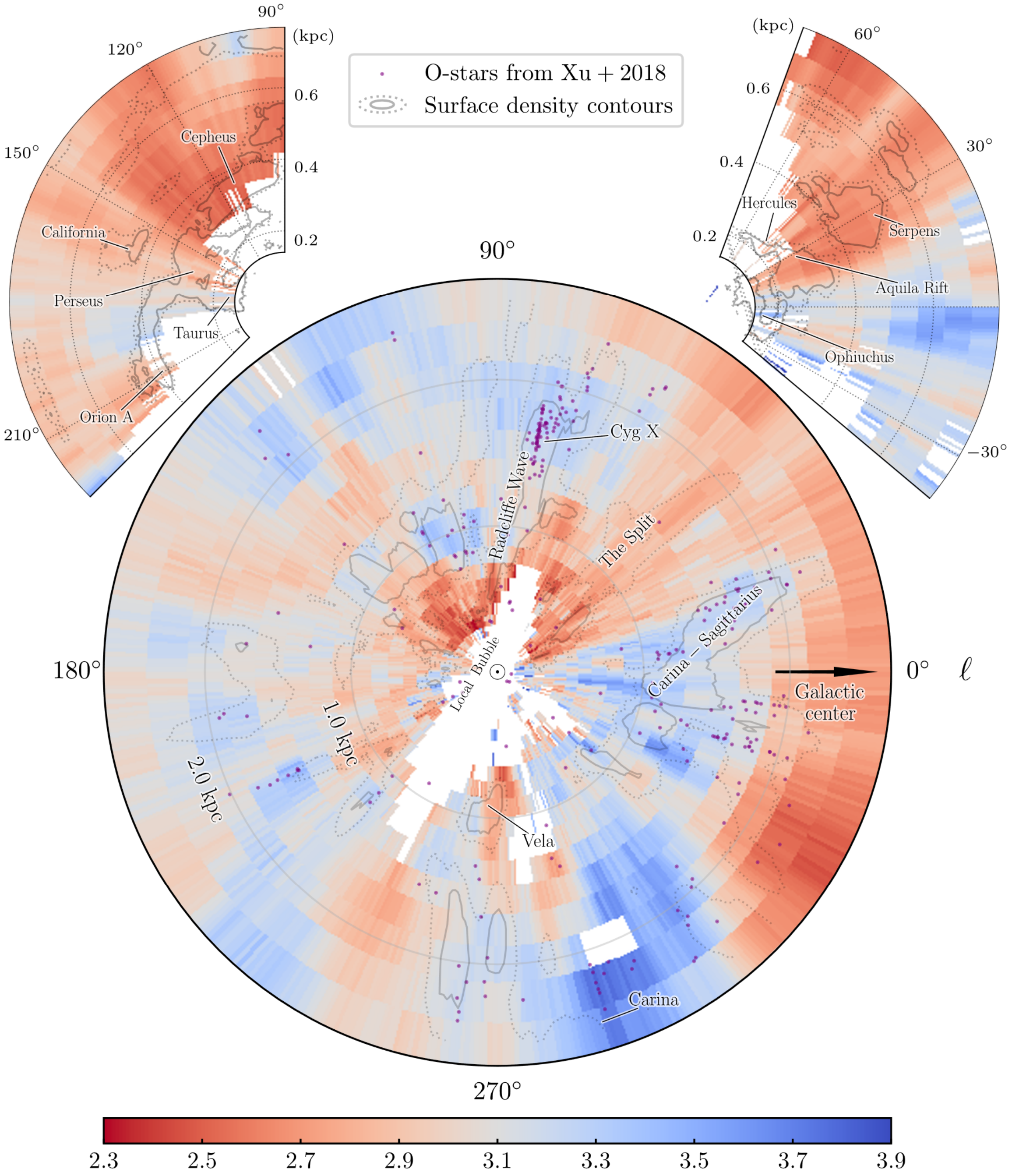

**Fig. 1. A bird's-eye view of $R(V)$ in the Galactic plane**, with the Sun located at the center. We calculate averaged $A(440\ \mathrm{nm})$ and $A(550\ \mathrm{nm})$ by integrating along the z-axis from $z = -400$ pc to $+400$ pc, and convert to averaged $R(V)$ by Eq. (B8). We select high-extinction ($E > 0.1$) stars that pass all of our observational and fitting quality cuts (See Appendix D.4 for details). High- (and higher-) extinction regions are enclosed by dashed (and solid) contours. Differences in $R(V)$ between high-density regions are correlated with the local number density of O/B-stars (denoted by purple dots, from Xu et al. *22*), indicating a possible link to star formation. For example, Carina-Sagittarius and Cygnus X host high densities of O/B-stars and have high $R(V)$, while the Split and the nearby regions of the Radcliffe Wave, which lack O/B

stars, have low $R(V)$. We are unable to measure $R(V)$ in the Local Bubble because of the low dust density in this region.

In Fig. 2, we show the sky distribution of average differential $R(V) \equiv \Delta A(V) / \Delta E(B-V)$ in distance slices. High- (and higher-) extinction regions are enclosed by dotted (and solid) contours, with positions of several dust clouds labeled according to Zucker et al. (*28*). Overall, dust clouds show lower-than-average $R(V)$, possibly indicating accretion of metals from the gas phase onto the dust in these environments (*23*), although there are exceptions due to complicated local environment. We also find lower-than-average $R(V)$ values towards the Galactic Center (in the $2-4$ kpc distance slice), similar to what has previously been found using photometry from OGLE (*24*).

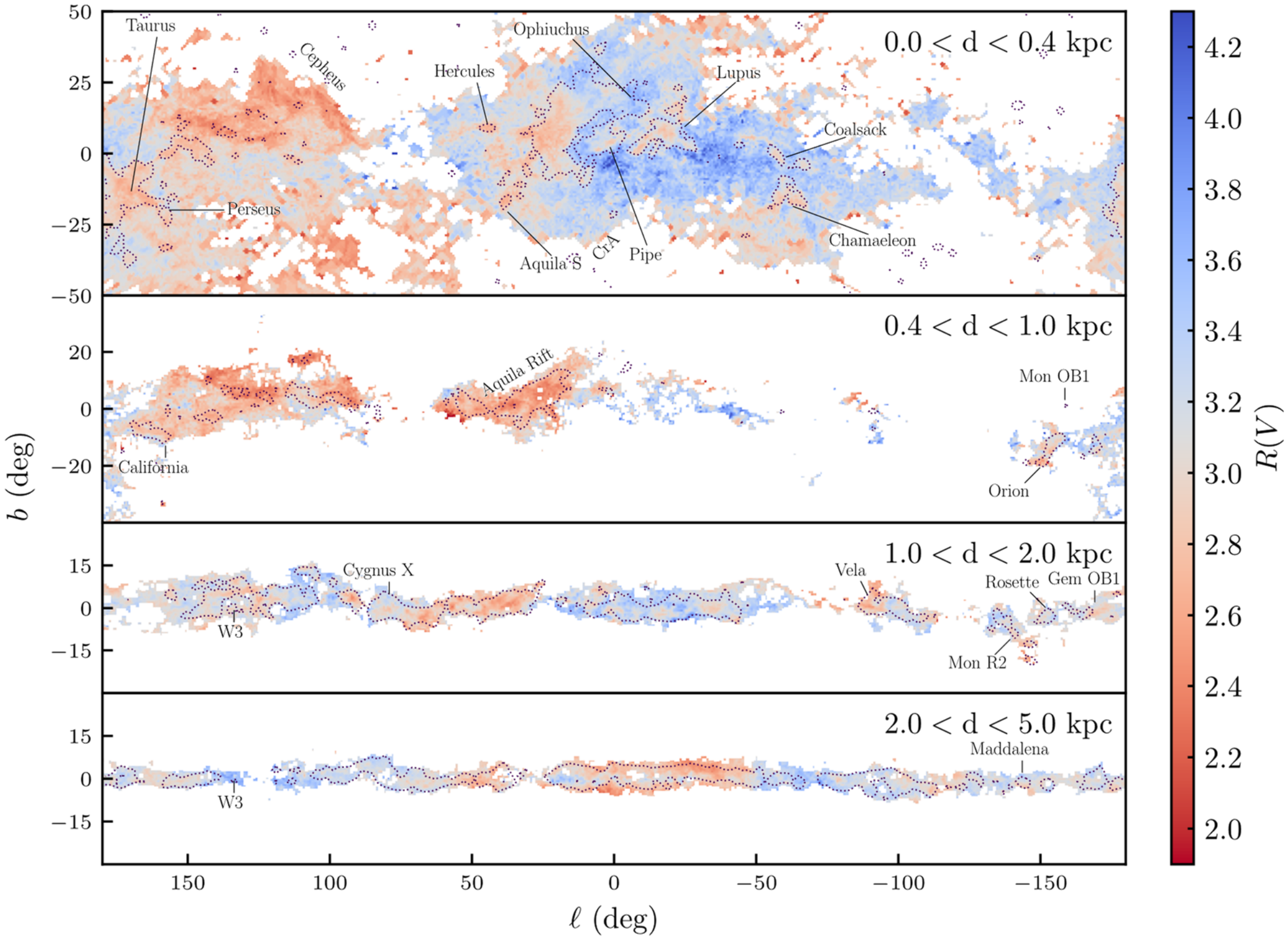

**Fig. 2. Sky distribution of R(V) in discrete distance bins.** We use the same selection standard as Fig. 1 to calculate differential $\Delta R(V) \equiv \Delta A(V) / \Delta E(B-V)$ for each line of sight in each distance bin. High-extinction regions are enclosed by purple dotted contours. We find kiloparsec-scale variation of $R(V)$, which is consistent with results based on a far smaller catalog published previously (*4*, *5*). However, we measure $R(V)$ for orders of magnitude more stars than ever before, with a much more complete sky coverage, allowing us to map variation in the extinction curve with greater spatial resolution and completeness. We find that $R(V)$ tends to decrease in intermediate-density regions of clouds, possibly indicating of accretion from the gas phase onto the dust in these environments (*23*), although some molecular clouds show complex patterns of $R(V)$. We also find lower-than-average $R(V)$ values towards the Galactic Center (in the 2–4 kpc distance bin), similar to what has previously been found using photometry from OGLE (*24*).

We also find $R(V)$ variation patterns in molecular clouds. In Fig. 3, we show zoomed-in views of five clouds, with high-extinction regions enclosed by solid contours. In this figure, to calculate average $R(V)$, we weigh the stars by extinction ($E$). That is, we calculate $\langle R(V)\rangle$ by

$$\langle R(V)\rangle = \langle ER(V)\rangle/\langle E\rangle. \quad [3]$$

We find that $R(V)$ is lower than average in the outer regions (also the intermediatedensity regions of dust) of many clouds, such as Orion A and Taurus and part of Mon OB1, which is consistent with what we find in Fig. 2. However, in the lines of sight towards central regions (or the densest regions) of the molecular clouds, we find far higher $R(V)$ values than in the outer regions. One possible explanation for this trend is that grains grow primarily by accretion in the translucent (intermediatedensity) medium, but that coagulation becomes the dominant grain-growth mechanism at higher densities. We discuss this possibility in greater detail in Section 4. The Perseus cloud shows a relatively complex pattern, which might indicate the complexity of the environment around it. In both the central and outer regions of $\rho$ Ophiuchus, $R(V)$ is higher than the average value in the Milky Way ($\sim$ 3.1), which may be related to the recent and ongoing star formation in this region, similar to the case of Carina-Sagittarius. In Fig. 4, we zoom in on the $\rho$ Ophiuchus and Upper Scorpius region, indicating the locations of O- and B-stars. This region hosts the nearest O/B-star associations, has been actively forming stars over the last ~20 million years, and may have been the site of a number of recent supernovae (*29–31*). We discuss a possible linkage between star formation and higher $R(V)$ values in Section 4.

We also obtain large numbers of $R(V)$ measurements in the Magellanic Clouds. In Fig. 5, we use $\sim 87{,}000$ stars in the Large and $\sim 8{,}900$ stars in the Small Magellanic Clouds (the "LMC" and "SMC" respectively; selected by imposing a cut of $\varpi + 3\sigma_{\varpi} < 1/15$ mas on the modeled parallax, and a "CSFD" (*32*) reddening cut of $> 0.1$, in addition to our standard cut on high-quality $R(V)$ measurements) to plot the inverse-variance-weighted mean $R(V)$ across both galaxies. We find a low average $R(V)$ (compared to the Milky Way) in both galaxies, as expected, though some star-forming regions in the LMC have higher $R(V)$.

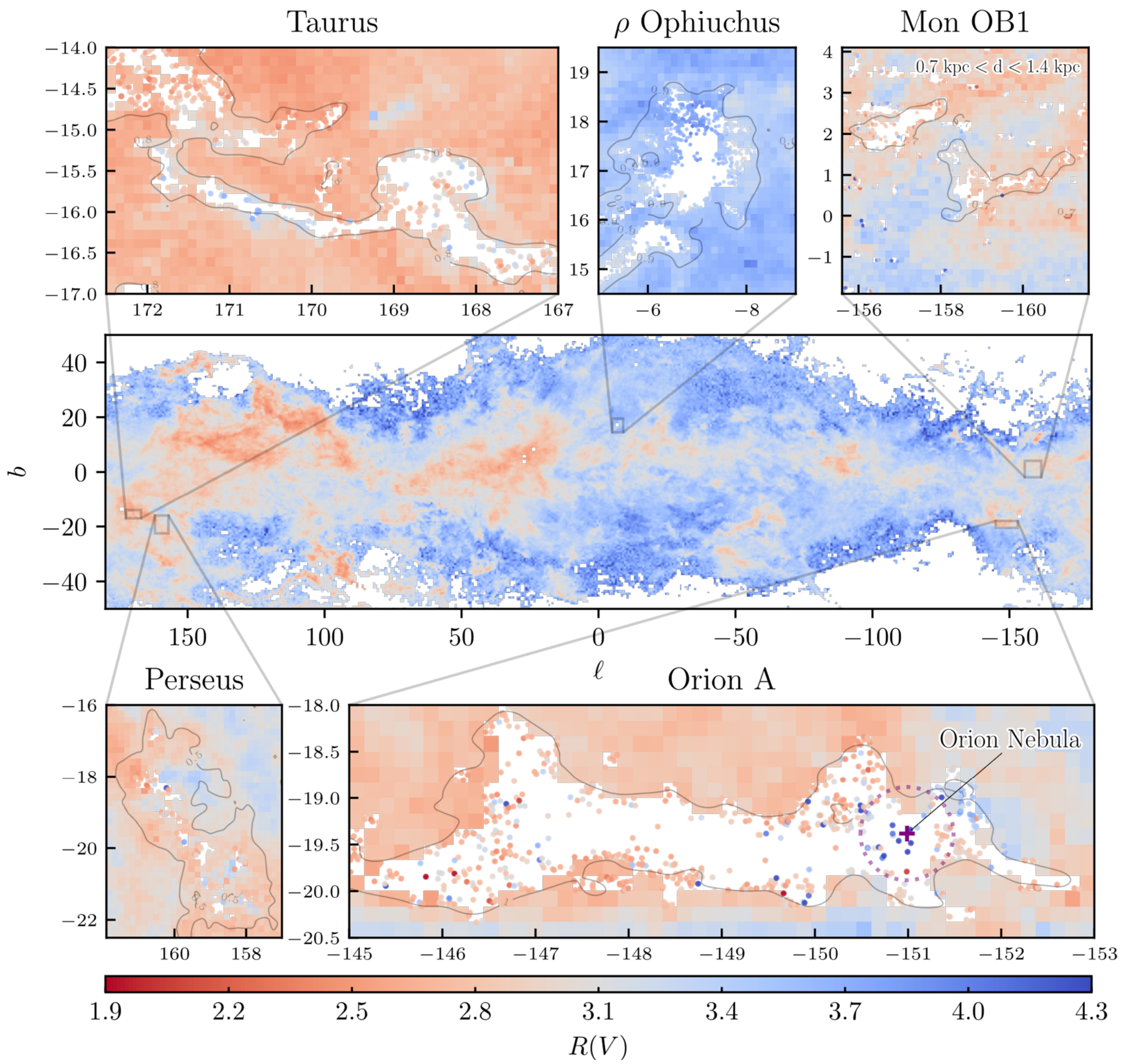

**Fig. 3**. **Averaged $R(V)$ distribution over the sky, and zoom-ins of five clouds.** We average over $R(V)$ measurements of stars that pass same the quality cuts as in Fig. 1. Gray contours enclose regions of high integrated extinction. In highly extincted regions with few observed stars, we show individual stars. We measure rich spatial structure in the extinction curve on many scales. $R(V)$ values in the outskirts of the Orion A molecular cloud are typically lower than the average value in the Milky Way ($\sim$ 3.1), possibly caused by accretion of metals from the gas phase onto the dust (*23*). However, the densest regions of the cloud, such as the Orion Nebula, have higher $R(V)$ than the outskirts, which is consistent with coagulation being the dominant mode of grain growth in these dense regions (*23*). A similar pattern also appears in Taurus. The clouds Mon OB1 and Perseus show complex spatial patterns in $R(V)$; their correlation with the local environment is worthy of further study. In the vicinity of $\rho$ Ophiuchus, both dense and diffuse regions show higher-than-average $R(V)$, indicating a higher relative abundance of large grains. Two possible explanations for the large grain sizes in $\rho$ Ophiuchus are recent dispersal of a dense molecular cloud by stellar feedback or destruction of small grains by recent supernova shocks. The anomalously high $R(V)$ of $\rho$ Ophiuchus has been noted by previous studies (*1*, *25*), but our 3D reconstruction of $R(V)$ (See Fig. 1) suggests that it may, in fact, be typical of regions with intense star formation.

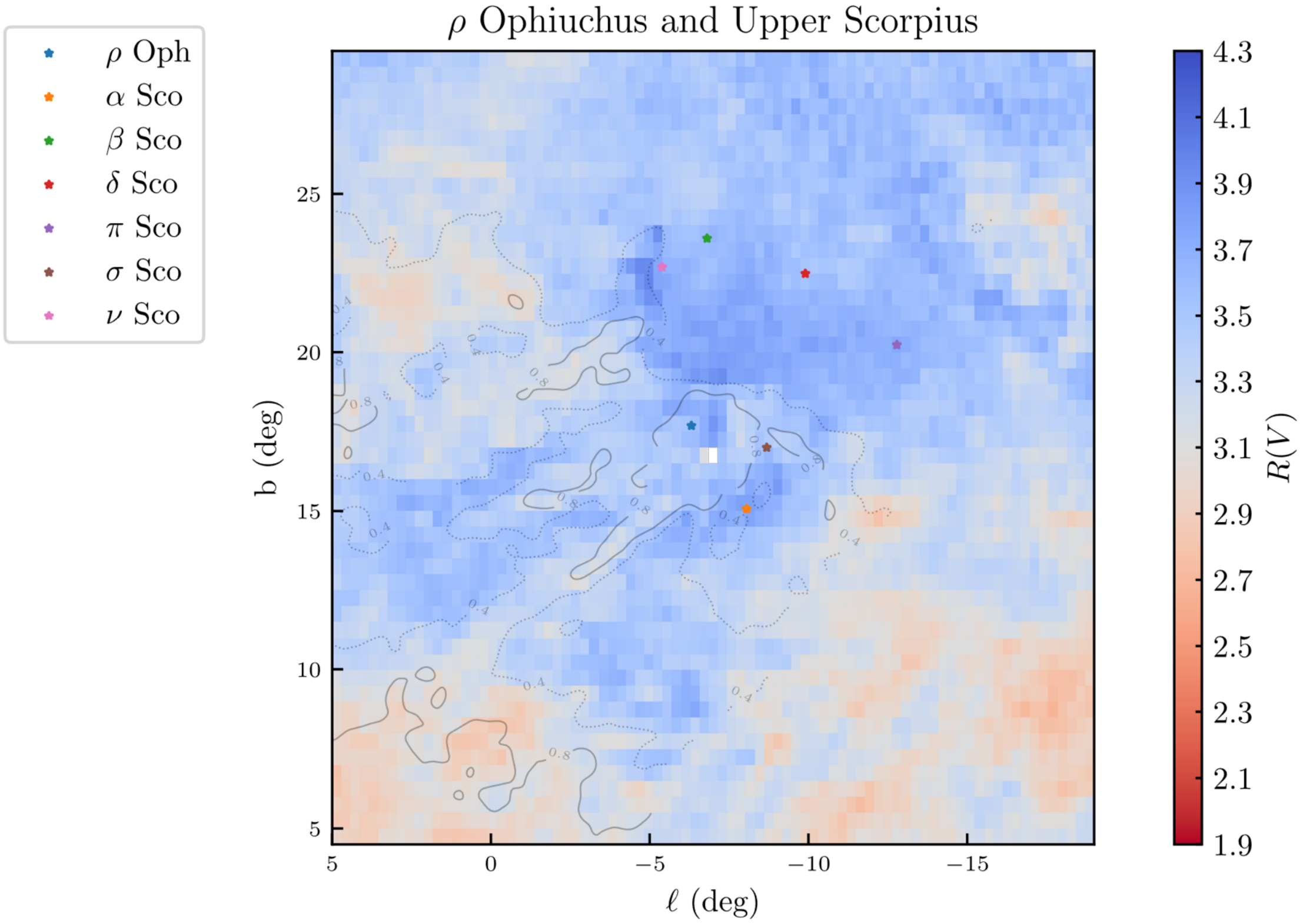

**Fig. 4. $R(V)$ of the nearby star-forming $\rho$ Ophiuchus and Upper Scorpius regions**, which host the nearest O/B-star associations to Earth. High-extinction (and higher-extinction) regions are enclosed by dotted (solid) contours. We additionally mark the positions of local O/B-stars. In 2D sky projections, this region stands out as having high $R(V)$ in both high- and low-dust-density regions. However, our 3D reconstruction of $R(V)$ in the Milky Way (See Fig. 1) reveals that more distant star-forming regions embedded in the Galactic disk also often have high $R(V)$. Due to projection effects, these more distant, embedded regions do not stand out in projected sky plots of $R(V)$.

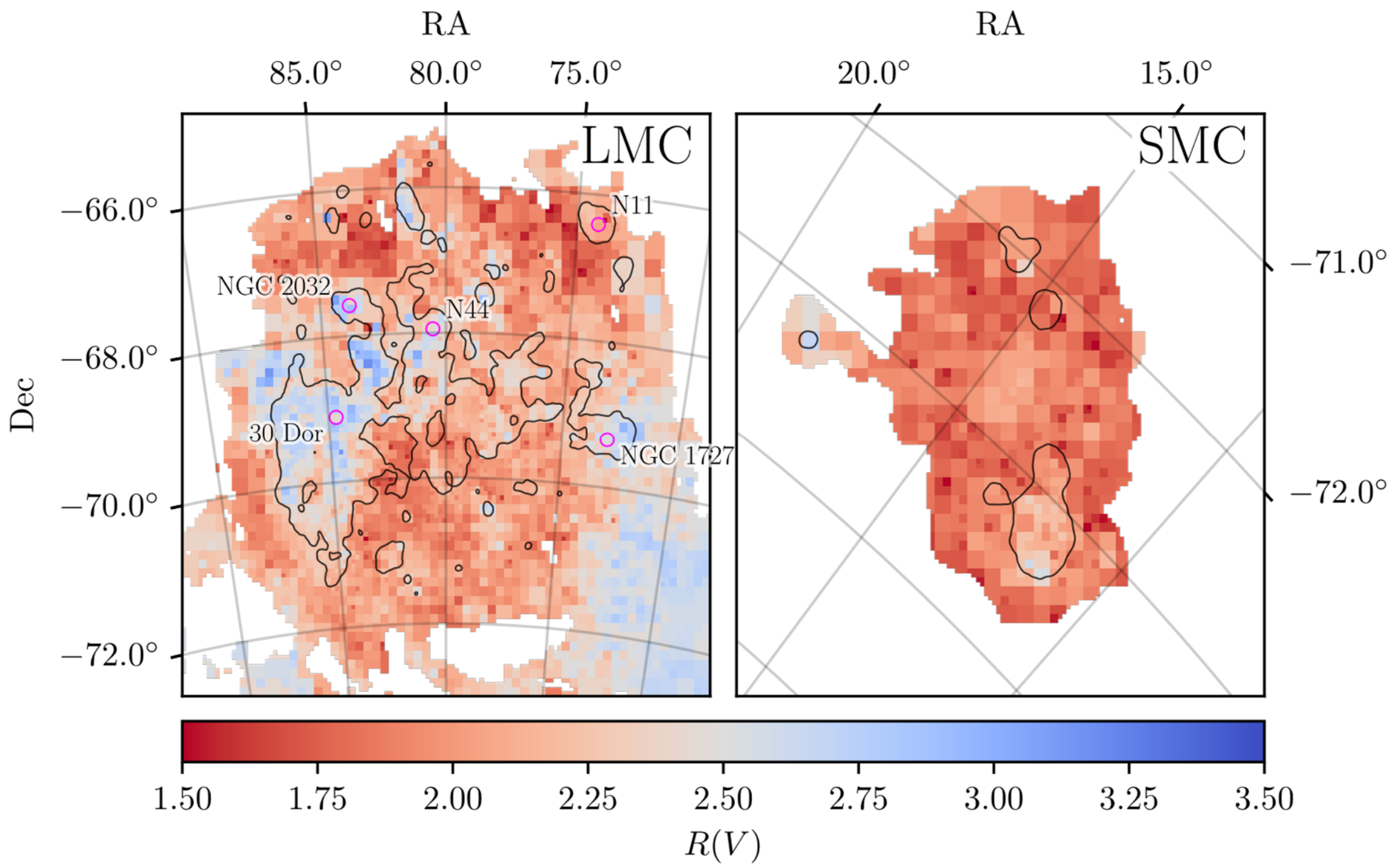

**Fig. 5. $R(V)$ in the Large and Small Magellanic Clouds** (the LMC and SMC, respectively). The solid curves are iso-contours with SFD reddening of 0.5, and are intended to indicate the general shape of the Magellanic Clouds. A number of star-forming regions in the LMC are indicated by cyan circles. Although the average $R(V)$ in both Magellanic Clouds is below the Milky-Way average, there are regions with higher $R(V)$, such as the Tarantula Nebula (30 Doradus), NGC 2032 and NGC 1727 star-forming regions.

## Discussion

A striking feature of our results, which can be seen in the central panel of Fig. 3, is that $R(V)$ tends to be higher than average in highly diffuse regions, and then tends to decrease as dust density increases. This is contrary to the common picture that $R(V)$ increases with density in the ISM, as grain-growth processes lead to a larger grain size. However, in some very dense regions, such as the Orion Nebula and dense cores in Taurus, we find that $R(V)$ sharply increases. Thus, in general, we observe decreasing $R(V)$ with dust density in the translucent (moderate-density) ISM, but high $R(V)$ in the densest inner regions of clouds. We propose a possible physical explanation for this U-shaped trend of $R(V)$ with density, based on the separate effects of accretion and coagulation on the dust grain size distribution.

The two dominant mechanisms of dust grain growth in the interstellar medium are accretion of elements from the gas phase onto the surface of grains, and coagulation of grains that collide and "stick" together. While both processes increase average grain size, their effects on the grain-size distribution – and thus $R(V)$ – are very different.

To a rough approximation, accretion tends to deposit a layer of approximately equal thickness onto each grain. The resulting fractional increase in surface area is therefore much larger for small grains than for large grains. Even though all grains increase in size, the total cross-section of small grains grows more quickly, and the cross-section-weighted average grain size *decreases*.

The contribution of the small grains to the extinction curve thus becomes greater, causing a steepening of the extinction curve, which is equivalent to decreasing $R(V)$.

Coagulation, by contrast, transforms pairs of small grains into larger grains. It therefore tilts the grain-size distribution, decreasing the abundance of small grains and increasing the abundance of large grains. In addition, the grain-size distribution develops a tail of large grains. As a result, the extinction curve becomes flatter, which is equivalent to increasing $R(V)$. For a more detailed (and numerical) discussion of the interplay between accretion and coagulation, see Hirashita (*23*).

As accretion involves interactions of dust grains with gas, while coagulation requires grain-grain interactions, coagulation occurs preferentially at higher densities. Our results are consistent with a picture in which in the less dense, "translucent" regions of interstellar clouds, accretion is the dominant mechanism of grain growth, leading to the observed decrease in $R(V)$ with dust density. However, in the densest regions of clouds, our results suggest that coagulation takes over, causing $R(V)$ to rapidly increase.

A number of lines of evidence suggest that recent star formation correlates with high $R(V)$. We observe high $R(V)$ across the entire Upper Scorpius and $\rho$ Ophiuchus region (See Fig. 4), which hosts the nearest O/B-star associations to the Sun, and which is a site of active and recent star formation (*29–31*). While this region stands out in 2D sky projections as having high $R(V)$ in both low- and high-dust-density regions, it may be typical of star-forming regions. A bird's-eye view of $R(V)$ in the Milky Way disk shows that concentrations of O/B stars tend to lie in regions of high $R(V)$. For example, though much of the Radcliffe Wave shows low $R(V)$ and is relatively devoid of O/B stars, the most distant region of the wave, Cygnus X, hosts a large concentration of O/B stars and has higher-than-average $R(V)$. The Split and Vela, which host few O/B stars, have low $R(V)$, while the Carina-Sagittarius arm, which hosts a higher density of O/B stars, has high $R(V)$. This correlation is also apparent in the LMC, where several star-forming regions (marked in Fig. 5) have relatively higher $R(V)$. There are a number of possible physical mechanisms for the correlation between star formation and high $R(V)$, among them the preferential destruction of small grains in supernova shocks (*33*), the formation of large grains in dense molecular cores (*34*, *35*), and the subsequent dispersal of these large grains by stellar feedback.

Finally, we remark briefly on the anti-correlation between the slope of the far-infrared thermal dust emission spectrum, parameterized by $\beta$ (*36*), and $R(V)$, which was originally found by S16 (*4*). We confirm S16's results along the low-Galacticlatitude sightlines that S16 probed. As noted in S16, these complex sightlines, with multiple clouds along the line of sight, are not ideal for determining whether $\beta$ and $R(V)$ are anti-correlated at a cloud-by-cloud level. We repeat S16's analysis for clouds at higher Galactic latitudes, and find a weaker – but still modest – anti-correlation between $\beta$ and $R(V)$. For more details, see Appendix E.

## Conclusions

We have determined $R(V)$ of the foreground dust for all stars with Gaia XP spectra. $\sim 130$ million of these measurements are of high quality. This enables mapping of the variation of the dust extinction curve in the Milky Way and Magellanic Clouds with both unprecedented scale and detail. To determine $R(V)$, we build an empirical forward model that maps stellar atmospheric parameters ($T_{\mathrm{eff}}$, $[\mathrm{Fe/H}]$, $\log g$), extinction column density ($E$), extinction curve variation ($\xi$), and parallax ($\varpi$) onto *Gaia* XP spectra and near-infrared bands from 2MASS (J, H, Ks) and WISE (W1, W2). In this model, the extinction curve varies within a one-parameter family (as in Cardelli et al. *1*, Fitzpatrick *2* and other extinction-curve models). The model parameters are directly learned from data, without imposing constraints from prior knowledge. We train this model using the crossmatch of XP spectra and the LAMOST DR8 catalog of atmospheric parameters. This model is built in an auto-differentiable framework, TensorFlow 2, which enables propagation of observational uncertainties into uncertainties on stellar parameters.

We apply this model to all 220 million Gaia XP spectra. 99% of the XP sources do not have LAMOST atmospheric parameters, so we impose a weak prior on stellar types, based on the distribution of our training data, in order to prevent the optimizer from reaching regions of parameter space that are not covered by our training set. We find good agreement between our $R(V)$ results and existing catalogs that use orthogonal data and methods, although our catalog of $R(V)$ contains more than an order of magnitude more stars than any previous high-precision catalog. Our catalog is publicly available at https://zenodo.org/doi/10.5281/zenodo.10719756, along with our trained model of XP spectra and extinction curves. This paper explores a preliminary 3D map of $R(V)$ of the Milky Way, revealing spatially complex patterns on many scales. We find a U-shaped relationship between $R(V)$ and dust column density ($E$) in molecular clouds, such as Orion A and Taurus (i.e., $R(V)$ tends to decrease with E, and then to increase after E reaches some threshold). A possible explanation is that as E increases, the dominant process of dust grain growth switches from accretion to coagulation. Accretion favors smaller grains due to their higher surface area per unit mass, while coagulation consumes small grains to form larger ones. We also notice a strong correlation between high-$R(V)$ and star formation, as traced by O/B-stars, in both the Galactic disk and LMC. The nearest example of such a region is $\rho$ Ophiuchus / Upper Scorpius. This region appears anomalous in projected 2D sky maps, as it has high $R(V)$ across both highand low-density areas. However, our 3D dust reconstruction reveals high $R(V)$ in more distant star-forming regions embedded in the Milky Way disk, making $\rho$ Ophiuchus / Upper Scorpius' apparent uniqueness simply a matter of it being nearby and above the Galactic plane. Possible explanations for the correlation between high $R(V)$ and star formation include preferential destruction of small grains by supernova shocks and cycling of large grains formed in molecular clouds back into the low-density ISM following the dispersal of the clouds.

Finally, with $\sim 130$ million precisely measured $R(V)$ values, our results build the foundation of the "next-generation" of dust maps of the Milky Way that model the distribution of $E$ and $R(V)$ simultaneously. This will allow more accurate observational extinction corrections and open up a new window into the chemistry and dust physics of the ISM.

E. del Pozo, M. Delbo, A. Delgado, H. E. Delgado, F. di Marco, P. Di Matteo, S. Diakite, E. Distefano, C. Dolding, S. Dos Anjos, P. Drazinos, J. Durán, Y. Dzigan, E. Ecale, B. Edvardsson, H. Enke, M. Erdmann, D. Escolar, M. Espina, N. W. Evans, G. Eynard Bontemps, C. Fabre, M. Fabrizio, S. Faigler, A. J. Falcão, M. Farràs Casas, F. Faye, L. Federici, G. Fedorets, J. Fernández-Hernández, P. Fernique, A. Fienga, F. Figueras, F. Filippi, K. Findeisen, A. Fonti, M. Fouesneau, E. Fraile, M. Fraser, J. Fuchs, R. Furnell, M. Gai, S. Galleti, L. Galluccio, D. Garabato, F. García-Sedano, P. Garé, A. Garofalo, N. Garralda, P. Gavras, J. Gerssen, R. Geyer, G. Gilmore, S. Girona, G. Giuffrida, M. Gomes, A. González-Marcos, J. González-Núñez, J. J. González-Vidal, M. Granvik, A. Guerrier, P. Guillout, J. Guiraud, A. Gúrpide, R. Gutiérrez-Sánchez, L. P. Guy, R. Haigron, D. Hatzidimitriou, M. Haywood, U. Heiter, A. Helmi, D. Hobbs, W. Hofmann, B. Holl, G. Holland, J. A. S. Hunt, A. Hypki, V. Icardi, M. Irwin, G. Jevardat de Fombelle, P. Jofré, P. G. Jonker, A. Jorissen, F. Julbe, A. Karampelas, A. Kochoska, R. Kohley, K. Kolenberg, E. Kontizas, S. E. Koposov, G. Kordopatis, P. Koubsky, A. Kowalczyk, A. Krone-Martins, M. Kudryashova, I. Kull, R. K. Bachchan, F. Lacoste-Seris, A. F. Lanza, J.-B. Lavigne, C. Le Poncin-Lafitte, Y. Lebreton, T. Lebzelter, S. Leccia, N. Leclerc, I. Lecoeur-Taibi, V. Lemaitre, H. Lenhardt, F. Leroux, S. Liao, E. Licata, H. E. P. Lindstrøm, T. A. Lister, E. Livanou, A. Lobel, W. Löffler, M. López, A. Lopez-Lozano, D. Lorenz, T. Loureiro, I. MacDonald, T. Magalhães Fernandes, S. Managau, R. G. Mann, G. Mantelet, O. Marchal, J. M. Marchant, M. Marconi, J. Marie, S. Marinoni, P. M. Marrese, G. Marschalkó, D. J. Marshall, J. M. Martín-Fleitas, M. Martino, N. Mary, G. Matijevič, T. Mazeh, P. J. McMillan, S. Messina, A. Mestre, D. Michalik, N. R. Millar, B. M. H. Miranda, D. Molina, R. Molinaro, M. Molinaro, L. Molnár, M. Moniez, P. Montegriffo, D. Monteiro, R. Mor, A. Mora, R. Morbidelli, T. Morel, S. Morgenthaler, T. Morley, D. Morris, A. F. Mulone, T. Muraveva, I. Musella, J. Narbonne, G. Nelemans, L. Nicastro, L. Noval, C. Ordénovic, J. Ordieres-Meré, P. Osborne, C. Pagani, I. Pagano, F. Pailler, H. Palacin, L. Palaversa, P. Parsons, T. Paulsen, M. Pecoraro, R. Pedrosa, H. Pentikäinen, J. Pereira, B. Pichon, A. M. Piersimoni, F.-X. Pineau, E. Plachy, G. Plum, E. Poujoulet, A. Prša, L. Pulone, S. Ragaini, S. Rago, N. Rambaux, M. Ramos-Lerate, P. Ranalli, G. Rauw, A. Read, S. Regibo, F. Renk, C. Reylé, R. A. Ribeiro, L. Rimoldini, V. Ripepi, A. Riva, G. Rixon, M. Roelens, M. Romero-Gómez, N. Rowell, F. Royer, A. Rudolph, L. Ruiz-Dern, G. Sadowski, T. Sagristà Sellés, J. Sahlmann, J. Salgado, E. Salguero, M. Sarasso, H. Savietto, A. Schnorhk, M. Schultheis, E. Sciacca, M. Segol, J. C. Segovia, D. Segransan, E. Serpell, I.-C. Shih, R. Smareglia, R. L. Smart, C. Smith, E. Solano, F. Solitro, R. Sordo, S. Soria Nieto, J. Souchay, A. Spagna, F. Spoto, U. Stampa, I. A. Steele, H. Steidelmüller, C. A. Stephenson, H. Stoev, F. F. Suess, M. Süveges, J. Surdej, L. Szabados, E. Szegedi-Elek, D. Tapiador, F. Taris, G. Tauran, M. B. Taylor, R. Teixeira, D. Terrett, B. Tingley, S. C. Trager, C. Turon, A. Ulla, E. Utrilla, G. Valentini, A. van Elteren, E. Van Hemelryck, M. van Leeuwen, M. Varadi, A. Vecchiato, J. Veljanoski, T. Via, D. Vicente, S. Vogt, H. Voss, V. Votruba, S. Voutsinas, G. Walmsley, M. Weiler, K. Weingrill, D. Werner, T. Wevers, G. Whitehead, Ł. Wyrzykowski, A. Yoldas, M. Žerjal, S. Zucker, C. Zurbach, T. Zwitter, A. Alecu, M. Allen, C. Allende Prieto, A. Amorim, G. Anglada-Escudé, V. Arsenijevic, S. Azaz, P. Balm, M. Beck, H.-H. Bernstein, L. Bigot, A. Bijaoui, C. Blasco, M. Bonfigli, G. Bono, S. Boudreault, A. Bressan, S. Brown, P.-M. Brunet, P. Bunclark, R. Buonanno, A. G. Butkevich, C. Carret, C. Carrion, L. Chemin, F. Chéreau, L. Corcione, E. Darmigny, K. S. de Boer, P. de Teodoro, P. T. de Zeeuw, C. Delle Luche, C. D. Domingues, P. Dubath, F. Fodor, B. Frézouls, A. Fries, D. Fustes, D. Fyfe, E. Gallardo, J. Gallegos, D. Gardiol, M. Gebran, A. Gomboc, A. Gómez, E. Grux, A. Gueguen, A. Heyrovsky, J. Hoar, G. Iannicola, Y. Isasi Parache, A.-M. Janotto, E. Joliet, A.

**Acknowledgments:**

**Non-author contributions:**

The authors acknowledge discussions with Aigen Li, Brandon Hensley and Thomas Henning on dust physics, and with Hans-Walter Rix on general aspects related to the work and manuscript.

**Funding:**

This work was funded through GG's Sofja Kowalewskaja Award, granted by the Alexander von Humboldt Foundtion.

**Author contributions:**

XZ and GG collaboratively developed the forward model, wrote the code base to implement it, and ran the model training and stellar parameter inference pipelines. XZ and GG made equal contributions to the analysis of the results, generation of the figures, and writing of the manuscript.

**Competing interests:** None.

**Data and materials availability:**

Our trained stellar model and parameter inferences are available at https://zenodo.org/doi/10.5281/zenodo.10719756. Our code base is available at https://github.com/Astrozxy/gaia_XP_forward_model. This work makes use of publicly available data from Gaia, 2MASS and WISE.

# Supplementary Materials for

## Unveiling the Milky Way dust extinction curve in 3D

Xiangyu Zhang[1*†], Gregory M. Green[1†]

Corresponding author: xzhang@mpia.de

†These authors contributed equally to this work.

**The PDF file includes:**

Supplementary Text
Figs. S1 to S8
Tables S1

**Supplementary Text**

Appendix A. Comparison of extinction curve models

As described in Section 2, our extinction curve varies within a one-parameter family that is learned directly from the XP spectra, without prior knowledge. Here, we compare our extinction curves with Gordon et al. (*3*, hereafter "G23"), which is based on higher-resolution spectroscopic data. We define a distance metric between different extinction curves, allowing them to be matched (despite different definitions of $R(V)$) and compared.

The left panel of Fig. A1 shows our learned family of extinction curves (colored by $R(V)$) in the XP spectral range, while the right panel shows the fractional difference (as a function of wavelength) between our extinction curves and the matching curves in the G23 family of curves. The two families of curves match to within $\sim 5\%$ between $392-950$ nm, and to within $\sim 7\%$ at the red end of the XP spectrum, from $950-992$ nm. This level of agreement over a wide range of $R(V)$ values is encouraging, given the weakness of the priors we placed on the extinction curve. Both the mean and variation of the extinction curve at every wavelength is allowed to vary independently, with only a weak continuity prior.

There are some subtleties involved when comparing two different families of extinction curves. Let us assume that we have two families of curves, which we will call B and C, each described by a single parameter, which is equivalent to $R(V)$. Let us say that we want to compare the curve produced by family B for a particular value of $R(V)$ to the most similar curve in family C. The simplest approach would be to choose the curve in family C that has the same $R(V)$. However, there is no guarantee that this is the most similar curve in family C over the entire wavelength range. We therefore want to define a distance measure that allows us to find the most similar curve in family C. This is a distance measure on the space of extinction curves. In addition, we want this distance measure to be invariant to the overall scaling of each extinction curve, as two curves with identical shape but different normalization represent dust with the same intrinsic properties.

We use the following scale-invariant distance measure to compare two extinction curves, $A(\lambda)$ and $A'(\lambda)$:

$$s^2 \equiv \min_{E'} \left\{ \frac{\int [A(\lambda) - E' A'(\lambda)]^2 \mathrm{d}\lambda}{\int A^2(\lambda) \mathrm{d}\lambda} \right\} = 1 - \frac{[\int A(\lambda) A'(\lambda) \mathrm{d}\lambda]^2}{[\int A^2(\lambda) \mathrm{d}\lambda][\int A'^2(\lambda) \mathrm{d}\lambda]} , \qquad \text{[A1]}$$

with the optimal E′ being given by

$$E' = \frac{\int A(\lambda) A'(\lambda) \mathrm{d}\lambda}{\int A'^2(\lambda) \mathrm{d}\lambda} . \qquad \text{[A2]}$$

As can be seen from the solution for $s^2$, this distance measure is manifestly invariant under exchange of $A$ and $A'$, or scaling of either function by a constant, and the measure is zero if $A = \alpha A'$ (where $\alpha > 0$ is an arbitrary scaling constant). As long as both $A$ and $A'$ are positive for all $\lambda$ (a condition we can safely assume for extinction curves), the measure is non-zero if $A \neq A'$ (again, up to an arbitrary scaling).

It can also be shown that this distance measure satisfies the triangle inequality, and is therefore a valid distance metric. Because we consider all extinction curves related by a positive scale factor to be equivalent, and because our distance measure is manifestly invariant under positive scalings of $A$ and $A'$, we can restrict our proof to extinction curves with a standard normalization without loss of generality: $\int A^2(\lambda)d\lambda = 1$ over a chosen range in $\lambda$. With this chosen normalization, our metric reduces to

$$s^2 = 1 - \left[\int A(\lambda)A'(\lambda)\mathrm{d}\lambda\right]^2 . \qquad \text{[A3]}$$

The standard Euclidean distance metric is given by

$$d_2^2 = \int [A(\lambda) - A'(\lambda)]^2 \mathrm{d}\lambda = 2\left[1 - \int A(\lambda)A'(\lambda)\mathrm{d}\lambda\right], \qquad \text{[A4]}$$

where we have used the normalization property in the second equality. For normalized extinction curves, our distance metric, $s$, can thus be expressed as a transformation of the Euclidean metric, $d_2$:

$$s = \left(d_2 - \frac{d_2^2}{4}\right)^{1/2} . \qquad [A5]$$

This transformation is increasing and concave for $0 \leq d_2 < 2$, which is the range of possible Euclidean distances for normalized extinction curves. Additionally, $s(d_2 = 0) = 0$. The transformation from $d_2$ to $s$ is thus metric-preserving (See Lemma 1 in *37*), and $s$ is a valid metric.

This distance measure (Eq. A1) is thus a fully fledged distance metric on the space of extinction curves (positive-valued functions of a real number, treating functions related by an overall positive scaling as equivalent). This metric allows us to match extinction curves between different families of curves.

Note that when using our distance metric to match extinction curves, the $R(V)$ value of the matching curves is not necessarily the same. In general, we will choose a curve with a desired $R(V)$ from family B, and then find the closest curve in family C.

Based on the matching method described above, we look into our curves for the most prominent structures, the "bumps" in Massa et al. (*38*) and Gordon et al. (*3*). We determine the "continuum" component by matching our curves with G23, and then subtracting the polynomial component of the matched G23 curve. In Fig. A2, we show the residuals of G23 (top panel) and our model (bottom). We very tentatively see excesses in the regions of the bumps at $\sim 437$ and $\sim 487$ nm, which are consistent with the first two bumps in G23, but the peak of the third bump is at $\sim$ 662 instead of 630 nm, as in G23. We also see fluctuations between 900 and 1000 nm, which are probably due to calibration "wiggles" at the edges of the Gaia XP spectra.

## Appendix B. Definition of $R(V)$

$R(V)$ was historically defined based on the extinction in B and V bands:

$$R(V) \equiv \frac{A(V)}{E(B-V)} = \frac{A(V)}{A(B) - A(V)} \, . \qquad \text{[B6]}$$

However, broad-band extinction depends not only on the wavelength-dependent dust extinction spectrum, $A(\lambda)$, but also on the source spectrum of the object being extincted and on the response curve of the photometric bands. Thus, stars of different temperatures sitting behind the exact same dust will have different $R(V)$ values, and different variants of the UBV photometric system will produce different $R(V)$ measurements for the exact same star.

In this work, when we speak of $R(V)$, we are primarily interested in the properties of the dust itself. We thus pick a definition of $R(V)$ that is independent of the properties of the particular star under observation. There are two ways to do this. The first is to define a "spectroscopic" measure of the dust extinction curve that depends only on $A(\lambda)$ at the approximate central wavelengths of the B and V bands:

$$R(55) \equiv \frac{A(550\,\mathrm{nm})}{A(440\,\mathrm{nm}) - A(550\,\mathrm{nm})} \, , \qquad \text{[B7]}$$

where $A(440\,\mathrm{nm})$ and $A(550\,\mathrm{nm})$ are the monochromatic extinction at 440 and 550 nm. The second way to define $R(V)$ consistently is to pick a standard source spectrum, such as a particular CALSPEC spectrum (*39*), and a standard set of B and V filter response curves, such as the Johnson bandpasses [*40*]. One can then calculate the theoretical $A(B)$ and $A(V)$ one would observe with the given dust extinction curve (in the limit of low extinction, as $R(V)$ depends on the dust column density, even holding the shape of the extinction curve constant), and calculate $R(V)$. There can be moderate differences between $R(55)$ and $R(V)$ as calculated for different reference stellar spectra and filter transmission functions. There are approximately linear transformations between the different definitions of $R(V)$, with both differences in slope and zero-point offset. In this work, unless otherwise noted, we use

$$\mathrm{R(V)} \approx 1.1\mathrm{R(55)} + 0.07, \qquad \text{[B8]}$$

as a proxy for $R(V)$, which we find to closely match the $R(V)$ one would obtain using the Johnson bandpasses and the CALSPEC Eta Ursa Majoris model spectrum.

## Appendix C. Correlation between $R(V)$ and $E$ in molecular clouds

In Section 3 and Fig. 3, we discuss patterns of $R(V)$ patterns in molecular clouds. In general, $R(V)$ tends to decrease in the regions of intermediate dust column density, and to increase in denser regions. In this appendix, we explore these trends quantitatively.

In Fig. C3, we show $E$ and $R(V)$ using sightlines towards molecular clouds. For each line of sight, we refer to the Bayestar19 reddening (*18*) behind the dust clouds as an indicator of density of dust, and calculate average $R(V)$ based on the stars in the line of sight that are embedded in or which lie behind the dust clouds. We find a U-shape correlation between $E$ and $R(V)$ in Orion A, Perseus and Taurus (*i.e.*, $R(V)$ first decreases with $E$ and then increases), which is consistent with the visual impression in Figs. 2 and 3. The correlation in $\rho$ Ophiuchus is less straightforward, possibly due to the effect of recent star formation on the grain-size distribution the low-density ISM.

Appendix D. Stellar model specification and training

As described in Section 2, we build a forward model that predicts XP spectra and near-infrared photometry as a function of stellar and dust parameters. This appendix provides a detailed description of the model structure and the regularization we use to ensure plausible results. We tabulate the detailed training process, and describe the reliability model that generates quality flags.

Appendix D.1 Model structure

Fig. D5 describes the structure of our model, which consists of an "intrinsic" part that predicts the de-reddened flux and a physical part that accounts for distance and extinction. The structure of the "intrinsic" part utilizes a 4-layer feed-forward neural network (with weights $\mathcal{W}$ and biases $b$) to map atmospheric parameters $\theta \equiv (T_{\rm eff}, [\mathrm{Fe/H}], \log g)$ to the natural logarithm of the de-reddened flux at 1 kpc, which is similar to v1 (*11*), with the sole structural difference being that we now concatenate the atmospheric parameters to each hidden layer. We account for distance by multiplying the flux by the square of the model parallax, $\varpi^2$. In contrast with v1, we allow the slope of the extinction curve to vary within a one-parameter family for each star. The mathematical structure of our model is given in Eqs. 1 and 2.

Appendix D.2 Regularization

In order to prevent overfitting and favor physically plausible models, we introduce a number of regularization terms into our loss function.

We would like the extinction model to be continuous in wavelength, and for it to default to a universal extinction curve (without variation) if the data itself contains no signal of variation. We therefore apply an L2 penalty to the components of the extinction curve variation vector ($\Delta \ln \vec{R}$), and to the difference between neighboring components in both the mean extinction curve vector ($\ln \vec{R}_0$) and the extinction curve variation vector (to encourage continuity). The continuity term is weighed by the discrete difference of the inverse of wavelength. Mathematically, we summarize all penalty terms on the extinction model by the following equation:

$$\frac{0.1}{N_\lambda} ||\Delta \ln \vec{R}\,||_2 + \frac{0.1 \times 550^{-2}}{N_\lambda - 1} \sum_{i=1}^{N_\lambda - 1} \frac{(\Delta \ln R_{i+1} - \Delta \ln R_i)^2 + (\ln R_{i+1} - \ln R_i)^2}{(1/\lambda_i - 1/\lambda_{i+1})^2}. \qquad \text{[D9]}$$

The first term is to break the scaling degeneracy between $\xi$ and $\Delta \ln \vec{R}$ due to the lack of prior knowledge on $\xi$. There is no need for a similar term on the mean extinction itself because the scalar amount of extinction ($E$) is constrained by Bayestar19 and SFD, and cannot vary in scale. The second and the third terms are to encourage continuity of the extinction curve. We weigh them by the inverse of discrete difference of inverse wavelength before calculating L2 norm, to account for the higher difference due to higher wavelength interval at the infrared bands. Finally, we give a relatively weak overall weighing, 0.1, to the total penalty term to prevent loss of information.

We would also like the output of the intrinsic flux model ($F_\lambda$) to vary smoothly as a function of stellar type ($\theta$). When training the model, we therefore include regularization terms on both the gradient and second derivatives of $F_\lambda$ w.r.t. $\theta$. During each training step, we calculate two quantities, D1 and D2 for each training source, related to the gradient and 2nd derivatives of $F_\lambda$, respectively:

$$D_1 \equiv \sum_{i,\lambda} \left( \frac{\partial F_\lambda}{\partial \theta_i} \right)^2 , \quad \text{[D10]}$$

$$D_2 \equiv \sum_{i,j} \left\{ \frac{\partial}{\partial \theta_j} \left[ \sum_{\lambda} \left( \frac{\partial F_\lambda}{\partial \theta_i} \right)^2 \right]^{1/2} \right\}^2 , \quad \text{[D11]}$$

where the sums over $i$ and $j$ are over the components of stellar type $\theta$, and the sum over $\theta$ is over the wavelength samples. Note that we do not use the more straightforward 2nd derivative term

$$\sum_{i,j,\lambda} \left( \frac{\partial^2 F_\lambda}{\partial \theta_i \, \partial \theta_j} \right)^2 \quad \text{[D12]}$$

because it is more computationally intensive to calculate than $D_2$. We then add the following model smoothness regularization term to the loss function, which penalizes the standard deviation of $D_1$ and the mean of $D_2$:

$$\mathrm{std}(\mathrm{D}_1) + \langle \mathrm{D}_2 \rangle \quad \text{[D13]}$$

The first term, encourages the gradients of $F_\lambda$ to be uniform across stellar-type space, $\theta$. The second term encourages small second derivatives (i.e., curvature) of $F_\lambda$ w.r.t. $\theta$. We could add penalties on yet higher derivatives of $F_\lambda$, but doing so would incur greater computational cost, and we find that penalizing the lowest two derivatives already produces smooth models. Finally, as in v1 of our model (*11*), we impose an L2 penalty on the weights in the $F_\lambda$ neural network: $\langle w_i^2 \rangle$, where $\{w_i\}$ is the collection of all weights in all neuralnetwork layers. This penalty further encourages simple models of the intrinsic stellar flux, $F_\lambda$, as a function of stellar type, $\theta$.

Appendix D.3 Training process

During the training process, our general approach is to initially hold the parameters for which we have reasonable initial guesses fixed, and to learn the parts of the model for which we have no good guesses. We then slowly allow more parameters to vary, eventually allowing all parameters to update in the last training phase. We divide up our training process into phases, which are summarized in Table S1.

In phases 1–4, we fix $\xi = 0$ for all sources (so that $\Delta \ln \vec{R}$ has no effect), thus learning only a universal extinction curve, similar to v1 of the model.

In phase 1, we train the stellar model M and the mean extinction curve $ln\,\vec{R}_0$, assuming the stellar parameters from LAMOST ($\theta$), Bayestar ($E$) and Gaia ($\varpi$) are correct. Because we assume the input parameters are correct, we only train with high-quality ("HQ") sources during this phase, as defined by the following criteria:

- $\sigma(T_{\mathrm{eff}}) < 200\ \mathrm{K}$
- $\sigma([\mathrm{Fe/H}]) < 0.2\mathrm{dex}$
- $\sigma(\log g) < 0.2\mathrm{dex}$
- $\hat{\varpi}/\sigma_{\varpi} > 10$
- $\sigma_E < 0.1$

We train for 512 epochs, with initial learning rate of $10^{-4}$, and reduce it by a factor of 2 every 64 epochs.

In phase 2, we update stellar parameters $(\theta, E, \varpi)$ of the HQ sources, holding the model (trained in phase 1) fixed. We train for 512 epochs, with an initial learning rate of $5 \times 10^{-5}$, which we reduce by a factor of 2 every 64 epochs.

In phase 3, we again use only HQ stars, and update stellar parameters, model parameters and the mean extinction curve for 1024 epochs. We use an initial learning rate of $2.5 \times 10^{-5}$, which we reduce learning rate by 2 every 256 epochs.

Before phase 4, we reject outliers using self-cleaning, as in v1, to remove stars large flux residuals or which deviate significantly from the LAMOST stellar type measurements. The standards of stars being accepted into the next phase are:

- $|T_{\mathrm{eff,est}} - T_{\mathrm{eff,obs}}|/\sigma_{T_{\mathrm{eff}}} < 4$
- $|\log g_{\mathrm{est}} - \log g_{\mathrm{obs}}|/\sigma_{\log g} < 4$
- $|[\mathrm{Fe/H}]_{\mathrm{est}} - [\mathrm{Fe/H}]_{\mathrm{obs}}|/\sigma_{[\mathrm{Fe/H}]} < 4$
- $|\varpi_{\mathrm{est}} - \varpi_{\mathrm{obs}}|/\sigma_{\varpi} < 4$
- $\chi^2\,/\,DOF < 10$.

We then continue training both stellar parameters and model parameters for 1024 epochs. We use an initial learning rate of $10^{-5}$, which we reduce by a factor of 2 every 256 epochs.
At the end of phase 4, we obtain a model similar to v1, with a universal extinction curve. In the remaining training phases, we gradually introduce extinction curve variation, and then finally train all parameters simultaneously

In phase 5, we make a rough initial guess of the variation of the extinction curve as a clipped linear function of wavelength ($\lambda$, in nm):

$$\Delta \ln \vec{\mathrm{R}}(\lambda)_{ini} \begin{cases} \dfrac{\lambda - 550}{600}, & \lambda \in [392, 992] \\ 0.5, & for \ \{J, H, K_s\} \\ 0, & for \ \{W1, W2\} \end{cases}$$

Such an initial guess is to represent the slope change of extinction curves with $\xi$. Holding the model fixed, we learn $E$, $\xi$ and $\varpi$ for HQ stars with $E > 0.1$ (as extinctioncurve variation is unmeasurable for $E \sim 0$). In this phase, we thus obtain a rough initial estimate of $\xi$ for each star, based on our initial guess of $\Delta \ln \vec{R}$. We train for 512 epochs, with initial learning rate of $5 \times 10^{-5}$ , which we reduce by a factor 2 every 128 epochs.

Before updating the extinction-curve variation ($\Delta \ln \vec{R}$ ), we first update the mean extinction curve ($\ln \vec{R}_0$) in phase 6 for 512 epochs, using only HQ stars with $E > 0.1$ . In this phase, we also set free $E$, $\xi$ and $\varpi$. We use an initial learning rate of $5 \times 10^{-5}$, which we reduce by a factor of 2 every 128 epochs.

In phase 7, we train the extinction-curve variation ($\Delta \ln \vec{R}$ ), setting free $E$, $\xi$ and $\varpi$. The initial learning rate is $10^{-4}$ for $\Delta \ln \vec{R}$ and $10^{-5}$ for the stellar parameters. We train both parameters of 1024 epochs and reduce the learning rates by a factor of 2 every 256 epochs. For each batch of sources, we first update the global model parameters, and then update the stellar parameters.
We have trained the "intrinsic" stellar model and extinction model separately in phases 1–4 and 5–8. In the following phases, we further enhance the model. In phase 9, we use HQ stars with $E > 0.1$ to update all parameters. We train for 1024 epochs with initial learning rates of $10^{-5}$ for both stellar and model parameters, which we reduce by a factor 2 every 256 epochs.

Through epoch 9, we have been training the model based only on high-quality stars based on observations (labelled as "HQ"). To make full use of the training set, we update stellar parameters of all stars in the training set and carry out a self-cleaning process to select all non-outliers for later phases.

In phase 10, we focus on stars with high uncertainties on the priors of extinction based on Bayestar19 (*18*). We optimize their extinction ($E$) for 768 epochs with initial learning rate of $10^{-2}$ and reduce the learning rate by 2 every 64 epochs. In phase 11, we update all stellar parameters for 512 epochs with initial learning rate of $5 \times 10^{-5}$ and reduce the learning rate every 128 epochs. A self-cleaning is then carried out based on the newly updated parameters. Phase 12 and 13 are the final stages, in which we further enhance the model based on the updated parameters and removal of outliers in the self-cleaning. We first freeze the slope of the extinction curve ($\Delta \ln \vec{R}$ ), and update the rest of the model parameters and all stellar parameters. We train 512 epochs with initial learning rates of $10^{-5}$ for both model and stellar parameters. We reduce learning rates every 128 epochs. In phase 13, we set free all parameters and train long enough (1024 epochs) with a relatively small initial learning rate ($10^{-6}$ ) for both model and stellar parameters to converge.

In all of the phases above, when updating model parameters, we weigh stars by the inverse of the prior of the stellar types to balance the distribution in the train set. We also weigh stars by the inverse of density of $\xi$ based on its histogram in phase 6-13. We summarize the training process in Table S1.

Appendix D.4 Training process

Similar to v1 (*11*), we also build quality cuts to estimate our confidence based on the quality of fitting and the coverage of stellar types in the training set. We recommend using the basic cut defined as:

$$\mathrm{chi2_opt}/61 < 2^1 \quad [D14]$$

$$\mathrm{ln_prior} > 7.43 \quad [D15]$$

$$|\mathrm{gaia_parallax\text{-}parallax_est}|/\mathrm{gaia_parallax_error} < 10 \quad [D16]$$

[1]We use a relatively looser cut ( $\mathrm{chi2_opt}/61 < 5$) when building 3D maps to allow more stars with reasonable quality.

We also train separate "confidence" cuts on $T_{\mathrm{eff}}$, $[\mathrm{Fe/H}]$ and $\log g$ based on the agreement of our parameters with LAMOST, using the same design of classifier as v1 (*11*). In this version, we adjust the input of the classifier and the standard of "good" stars. We use the following features to predict the confidence:

1. ln_dplx2
2. asinh_plx_snr
3. asinh_g_snr
4. asinh_bp_snr
5. asinh_rp_snr
6. ln_phot_bp_rp_excess_factor
7. ln_ruwe
8. fidelity_v2
9. norm_dg
10. asinh_flux_chi_X; $X: 0-65$

ln_dplx2 is $\ln(\varpi_{\mathrm{est}} - \varpi_{\mathrm{obs}})^2$. No. 2-5 of the features are the inverse hyperbolic sine (asinh) of the signal-to-noise ratio of Gaia parallax, G band, BP band and RP band photometry. No. 6-7 are the natural logarithm (ln) of Gaia features phot_bp_rp_excess_factor and ruwe, and No. 8-9 are Gaia features. No. 10 of features are the inverse hyperbolic sine (asinh) of the normalized residuals of flux at all 66 different wavelengths, labelled by "X".

Compared with v1, we remove features of fitting $\chi^2$, predicted stellar parameters and their probability in the prior, because these features are already taken into account in the basic cut. We also remove features of the quality of Gaia photometric bands, such as phot_g_mean_mag and ln_bp_chi_squared, in order to avoid trivial solutions.

We still use the "floored $\chi^2$" as standard of "good" stars:

$$\chi_\epsilon^2 \equiv \frac{(\Delta x)^2}{\sigma_{\rm opt}^2 + \sigma_{\rm LAMOST}^2 + \epsilon^2}, \qquad [D17]$$

where $\sigma_{\rm opt}$ and $\sigma_{\rm LAMOST}$ are the uncertainties of our estimates and the LAMOST catalog. $\epsilon$ is an uncertainty floor, with $\epsilon = 100$ K for $T_{\rm eff}$, 0.1 dex for $[{\rm Fe/H}]$ and 0.1 dex for $\log g$. We define $\chi_\epsilon^2 > 4$ as bad stars, and $\chi_\epsilon^2 < 1$ as good ones. To balance the training set, we re-sample "good" and "bad" stars by:

• Downsampling the "bad" stars if the frequency of their stellar parameters is higher than 80% of all parameters in the total distribution; (Frequencies and the total distribution are estimated from bin counts.)
• Re-sample all the "good" stars so that their frequency is always $5\times$ the number of bad stars within the same bin.

We train the classifier with $80\%$ of the "good" and "bad" stars using the same procedures as v1, and validate the result by the rest of the $20\%$ stars. Our classifier assigns a number between 0 and 1 to each of the 220 million stars, with 1 begin complete confidence and 0 being no confidence. We use 0.5 as our confidence threshold (i.e., a stellar parameter of confidence $> 0.5$ is considered trustworthy).

We show the validation of our parameters by comparing them with relatively higher-resolution spectroscopic surveys. We compare our results with LAMOST using the validation set which is unknown to the model during the training. We also compare $[{\rm Fe/H}]$ with APOGEE. We show these comparisons in Fig. D4. Our result agrees with LAMOST and APOGEE no worse than v1. When building 3D maps of $R(V)$, we select stars that pass all following cuts:

- basic_cut
- $T_{\rm eff}$ confidence cut
- $E > 0.1$
- $\xi_{\rm err} < 0.2$

The basic_cut is an indicator the quality of fitting. $T_{\rm eff}$ confidence cut ensures reliable determination of the slope of the spectra, which is a necessary condition for $R(V)$ determination. The cut $E > 0.1$ removes low extinction stars of which $R(V)$ is arbitrary, and the cut $\xi_{\rm err} < 0.2$ removes stars with highly uncertain $R(V)$. $\sim 130$ million stars pass all these cuts simultaneously.

## Appendix E. Comparisons with previous $R(V)$ measurements

There are a number of previous catalogs of dust $R(V)$ for lines of sight to individual stars. Here, we compare to four previous catalogs: Schlafly et al. (*4*, *5*, hereafter "S16"), Zhang et al. (*6*, no relation; hereafter "ZYC23"), (*41*, hereafter "F22"), and (*3*, hereafter "G23"). In these comparisons, we apply our basic quality cut to our catalog, and additionally require that $E > 0.1$ (as it is impossible to measure the extinction curve when $E \sim 0$), $\sigma_\xi < 0.1$, and $T_{\rm eff}$ confidence $> 0.5$ .

S16 determined dust $R(V)$ for 37,000 stars using a combination of APOGEE spectroscopy (*42*) and optical-NIR photometry from Pan-STARRS 1 (*43*), 2MASS (*20*) and WISE (*44*, *45*). S16 expressed the photometric reddening vector as a linear combination of a mean reddening vector and a second component, representing variation about the mean. S16 measured the coefficient, termed x, of this variation component for each star. This coefficient may be converted to $R(V)$ using color transformations between Pan-STARRS 1 bands and B and V. However, we can more directly compare our extinction-curve measurements to those of S16 by directly comparing S16's $x$ parameter to our $\xi$. If there is a one-to-one mapping between $x$ and $\xi$, then our two measures are fundamentally measuring the same phenomenon. We calculate two different types of rank correlations, which measure whether the relation between two variables is monotonic: the Spearman rank coefficient, $r_s$, and the Kendall rank coefficient, $\tau$. Both are defined so that 0 indicates no correlation, while 1 indicates perfect positive correlation. As can be seen in the left panel of Fig. E6, we indeed find a strong correlation between $x$ and $\xi$ ($r_s = 0.816$, $\tau = 0.628$), using a set of 7,523 stars with high-quality measurements in both catalogs (as defined by the cuts described at the beginning of this section, and by bits 0, 1, 3, and 4 of S16's flags being unset).

S16 found that for APOGEE sightlines in the Galactic plane, $R(V)$ – as measured by stellar extinction – is anti-correlated with the slope of dust thermal emission in the far-infrared (FIR), $\beta$, as measured Planck Collaboration et al. (*36*). If this anticorrelation holds generally, it indicates that the same processes that modify the shape of the optical extinction curve also modify the efficiency of dust thermal radiation at FIR wavelengths. Here, we follow-up on S16's findings. We verify that for the sightlines probed by S16, our $R(V)$ measure is strongly anti-correlated with $\beta$ (left panel of Fig. E7). However, as S16 noted, these Galactic-plane sightlines are sub-optimal, as the stars do not necessarily probe the entire column of radiating dust seen by Planck in the FIR, and because Planck's assumption of a single modified blackbody breaks down when there are multiple clouds along one sightline. In the right panel of Fig. E7, we therefore compare $R(V)$ and $\beta$ along high-Galactic-latitude sightlines. We select stars with $|b| > 10$ deg, which are additionally at a minimum height of 400 pc above or below the Galactic midplane (using their $5\sigma$ uncertainties in estimated distance), which have a "corrected" SFD reddening (*32*) of greater than 0.05, and which fulfill our other criteria for $R(V)$ comparisons. We additionally cut out sightlines within 10 deg of the Large Magellanic Cloud, 5 deg of the Small Magellanic Cloud, and 3 deg of M31. Using only these high-Galactic-latitude sightlines, we find a smaller – but still moderate – anti-correlation ($r_s = -0.356$ for high Galactic latitudes, vs. $-0.665$ for S16 sightlines). In both comparisons, we average both $\beta$ and $R(V)$ measurements for stars on a HEALPix nside = 64 scale (~ 55 arcmin), and ignore sightlines with fewer than 16 stars.

ZYC23 determined dust $R(V)$ for ~3 million stars using a combination of LAMOST spectroscopy and near-UV, optical, and NIR photometry from SDSS (*46*), Gaia, Pan-STARRS 1, 2MASS, WISE. We compare our extinction-curve measurements to those of ZYC23 using 253,057 stars with high-quality measurements in both catalogs. For ZYC23, we require that their reliable flag be set to 1, and that their reported uncertainty in $R(V)$ be less than 0.05. We convert our own $\xi$ estimates to $R(V)$ using the transformation described in Appendix B. As shown in the middle panel of Fig. E6, our $R(V)$ measurements correlate strongly with those of ZYC23 ($r_s = 0.738$, $\tau = 0.553$).

We also compare the 2D distribution as the Figure 6 in ZYC23 in Fig. E8. Since most of the stars in ZYC23 have Gaia XP spectra, we select the common high-extinction ($E > 0.1$) stars that pass our basic cut, our $T_{\mathrm{eff}}$ confidence cut and have $R(V)_{\mathrm{err}} < 0.1$ in the ZYC23 catalog. We show the 2D distribution in Fig. E8. Patterns of $R(V)$ in both panels are consistent, i.e., $R(V)$ is high (low) in the top panel where the same region shows high (low) $R(V)$ in the bottom panel. This is not surprising since there is a monotonically increasing (but not identical) linear relation between $R(V)$ in ZYC23 and this work, as shown in Fig. E6.

F22 determined an $R(V)$-like parameter, $R_0$, for the dust towards ~123 million stars using optical-NIR photometry from *Gaia*, 2MASS, and WISE, as well as parallax measurements from Gaia. In the right panel of Fig. E6, we compare $1\%$ of our high-quality $R(V)$ determinations (converted from $\xi$ as in the preceding paragraph) with the best-fit $R_0$ determined by F22. We find a weak positive correlation ($r_s = 0.281,\ \tau = 0.189$). Given our much stronger correlations with the results from S16 and ZYC23, both of which use spectroscopic determinations of stellar properties, we conclude that F22, which relies only on broadband photometry and parallaxes, produces a far noisier estimate of the extinction curve slope.

G23 modeled the $R(V)$-dependent extinction curve from the UV through the midinfrared, using a compilation of high-resolution O/B-star spectra from Gordon et al. (*47*), Fitzpatrick et al. (*48*), Gordon et al. (*49*), and Decleir et al. (*50*). Because our model is trained using the LAMOST catalog, it performs poorly for O/B stars, and most stars of these classes fail our quality cuts. As a result, we cannot directly compare our $R(V)$ estimates to those of G23 on a star-by-star basis. Instead, we estimate the *R(V)* of each star in the G23 catalog using XP stars in the same vicinity (in 3D). As this requires a precise parallax measurement for each G23 star, we only stars in the G23 catalog with precise (parallax_over_error $> 10$) and reliable (fidelity_v2 $> 0.5$; See Rybizki et al. (*51*)) Gaia parallaxes. For each G23 star, we select XP stars within a 10 arcmin radius, with $\mathrm{parallax_over_error} > 10$, and with an estimated distance that is either between 90% and 120% of the G23 star's parallax distance or between 0 and 100 pc more distant than the G23 star. We then calculate the inverse-variance weighted $\xi$ of these nearby XP stars (applying a floor of 0.01 on $\sigma_\xi$), and convert the result to $R(V)$. This is our proxy for the $R(V)$ of the given G23 star. This process will work well if the nearby XP stars probe a similar dust column as the given O/B star. However, the O/B stars in the G23 catalog often reside in highly extincted regions with a complicated dust distribution, meaning that nearby stars are not necessarily a good proxy for any given G23 star. To mitigate this issue, we exclude G23 stars for which the average SFD reddening of the XP neighbors is less than 20% of the SFD reddening of the G23 star, for which there are fewer than four neighbors, for which the standard deviation of our $R(V)$ estimates for the neighbors is greater than 0.5, or for which the G23 catalog uncertainty on $R(V)$ is greater than 0.5. We obtain comparisons with 100 stars from the G23 catalog, which we plot in the right panel of Fig. E6, together with their corresponding uncertainties. We find a moderately strong correlation of $r_s = 0.601$ and $\tau = 0.437$ between our $R(V)$ estimates and those of G23. We find that for the G23 stars with the highest $R(V)$, our catalog tends to underestimate $R(V)$. One possible explanation for this effect is that the highest $R(V)$ values occur in highly localized environments, which are not effectively probed by neighboring XP stars.

It is worth noting that S16, ZYC23, G23 and our work rely on different spectroscopic data (APOGEE, LAMOST, IUE/STIS/Spitzer and Gaia XP, respectively), though all three works use NIR photometry from 2MASS and WISE. Given the different underlying spectroscopic data and methods, the good agreement between S16, ZYC23, G23 and our work gives us high confidence in the robustness of our results. The one $R(V)$ catalog that does not correlate strongly with our results, F22, relies solely on photometric and parallax measurements. Precise stellar temperature inferences are required to isolate the effect of extinction on stellar colors. Based on our results, Gaia XP spectra, at a resolution of $\mathrm{R} \sim 40 - 100$, enable sufficiently precise stellar temperature estimates, while broadband photometry alone (or only in combination with parallax) appears to be insufficient to precisely determine $R(V)$.

**Fig. S1. (A1)**

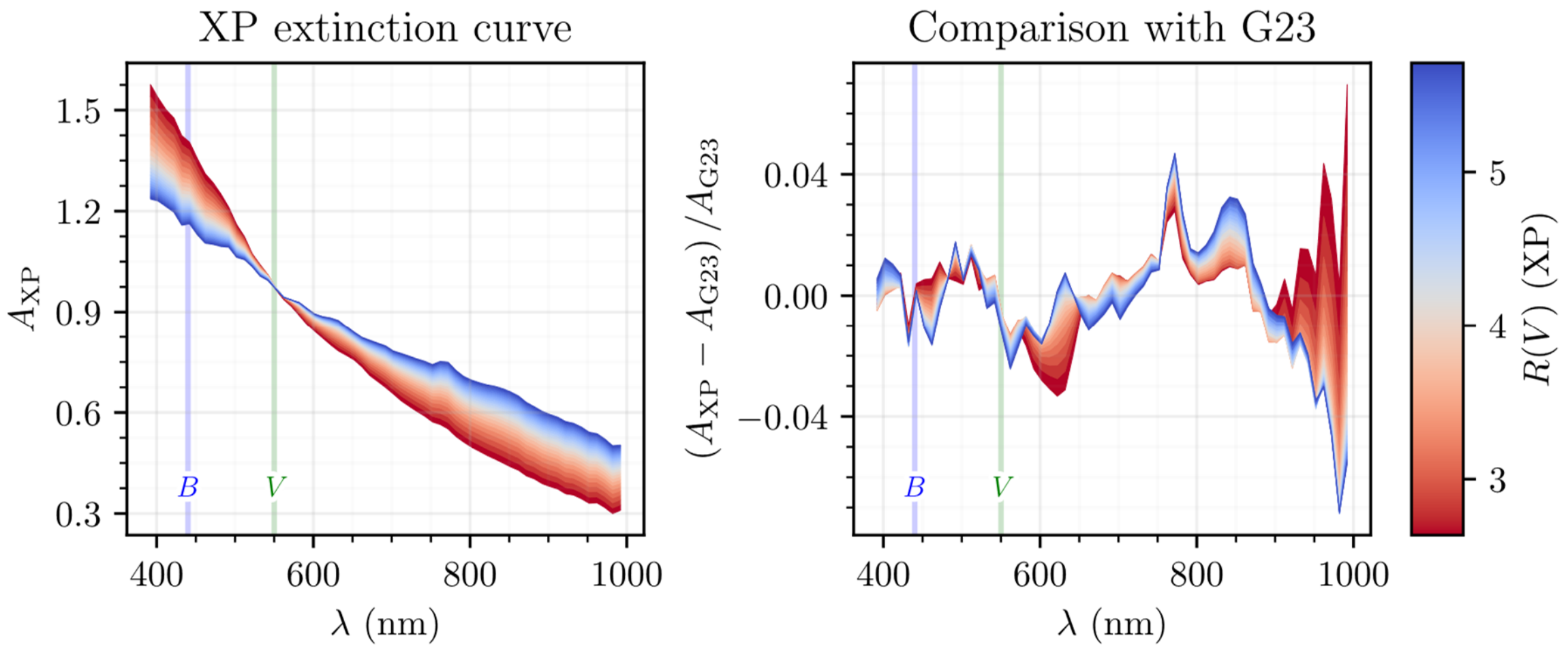

Fig. A1 Our extinction curve model (in the XP spectral range) as a function of $R(V)$ (left), and a comparison between our model and that of Gordon et al. (*3*, "G23"). In both panels, we mark the approximate central wavelengths of B and V bands, illustrating the insufficiency of individual wavelengths in describing the entire optical extinction curve. We find good agreement between our one-parameter family of extinction curves and that of G23, despite the fact that our model does not a priori know about existing extinction curve models and fits the extinction curve at each wavelength nearly independently.

**Fig. S2. (A2)**

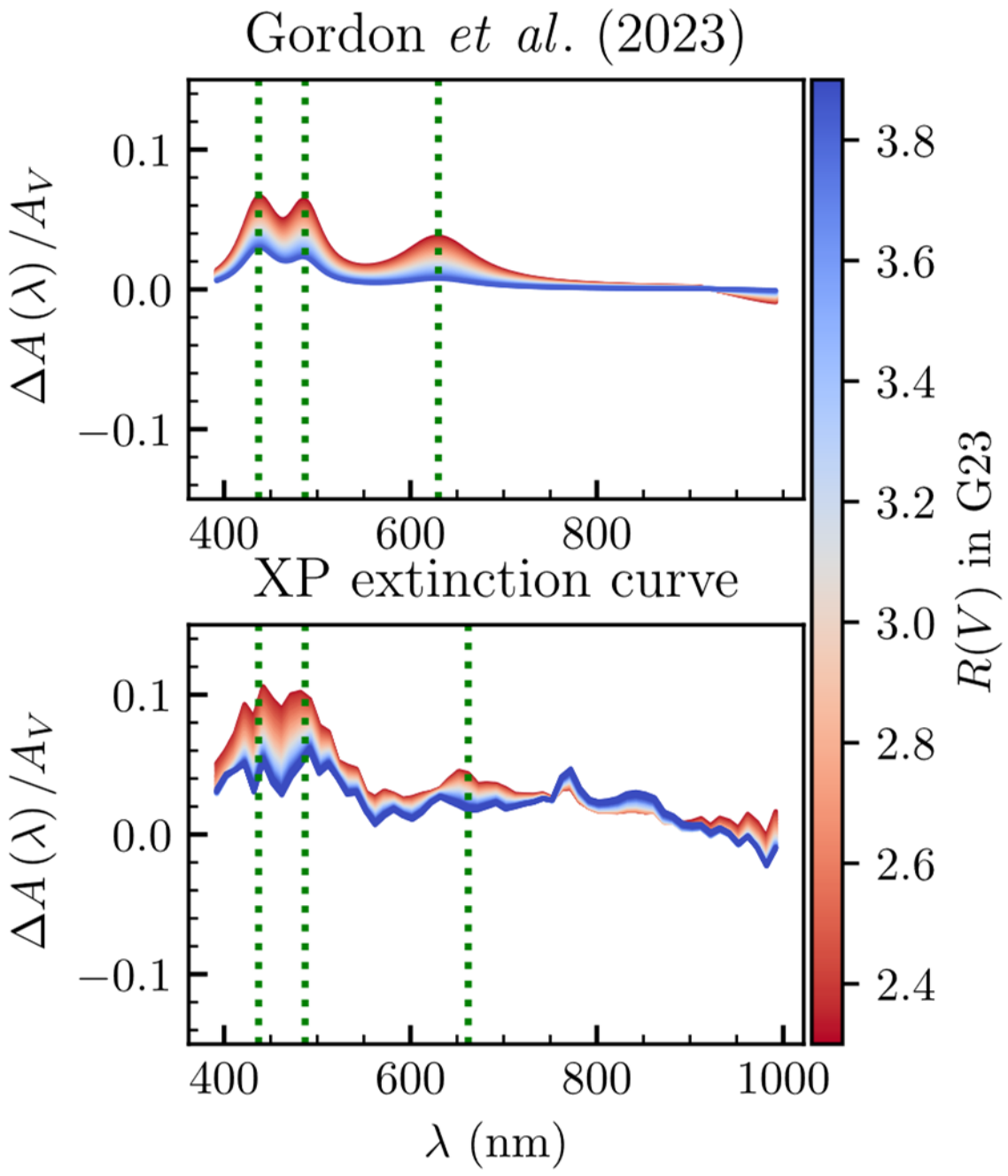

Fig. A2 "Bumps" in Gordon et al. (*3*) and in this work. The top panel shows the "bump" component in G23 (*3*), and the bottom panel shows the extinction curve in this work with the corresponding G23 polynomial component subtracted. The $R(V)$ value that determines the polynomial component is from the best match between G23's extinction curves and ours. We very tentatively see excesses in the regions of the bumps at $\sim 437$ and $\sim 487$ nm, which are consistent with the first two bumps in G23, but the peak of the third bump is at $\sim 662$ nm, instead of $\sim 630$ nm, as in G23. We also see fluctuations between $900$ and $1000$ nm, which are probably due to calibration "wiggles" at the edges of the XP spectra.

**Fig. S3. (C3)**

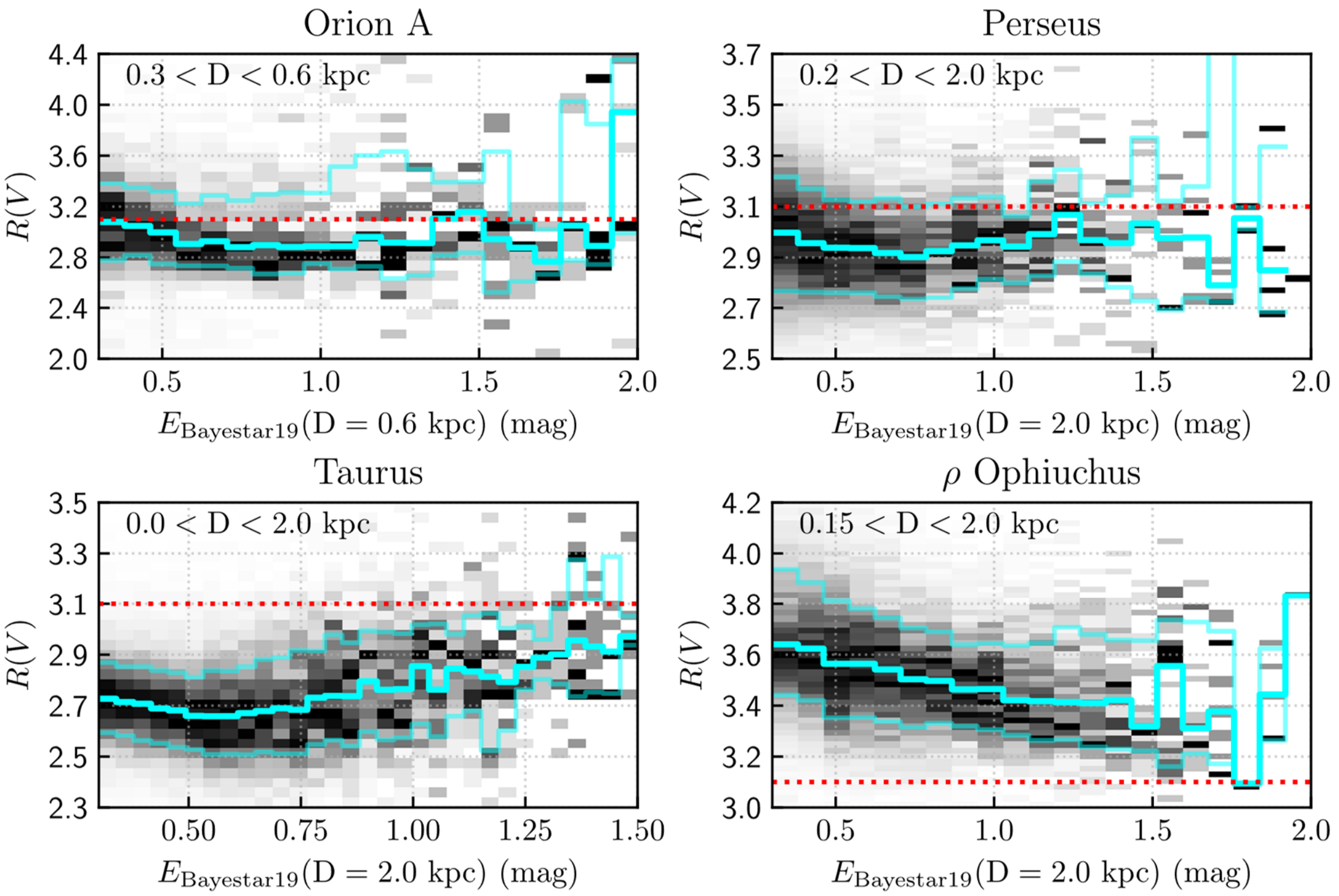

Fig. C3 Correlation between $R(V)$ and dust density for different lines of sight towards molecular clouds Orion A, Perseus, Taurus and $\rho$ Ophiuchus. The average $R(V)$ in the Milky Way (~ 3.1) is labeled by a red dotted line in each panel. Due to the selection function of Gaia XP, stars behind dense regions of molecular cloud are in lack of spectroscopic data, but $R(V)$ values of the lines of sight towards these regions can be determined by stars embedded onto the clouds. Therefore, we use the extinction value behind the clouds (as noted in each panel) from the 3D Bayestar19 map (*18*) ($E_{\mathrm{Bayestar19}}$) to represent the column density. Generally, Taurus and $\rho$ Ophiuchus have very different $R(V)$ values from the average case in the Milky Way, while Orion A and Perseus are closer to the average. Orion A, Perseus and Taurus show U-shaped curves: $R(V)$ tends to decrease with $E$ at low extinctions, but begins to increase at high $E$. The $\rho$ Ophiuchus region shows a relatively complex correlation between $R(V)$ and $E$, with no obvious turning point, which may be related to recent and ongoing star formation.

**Fig. S4. (D4)**

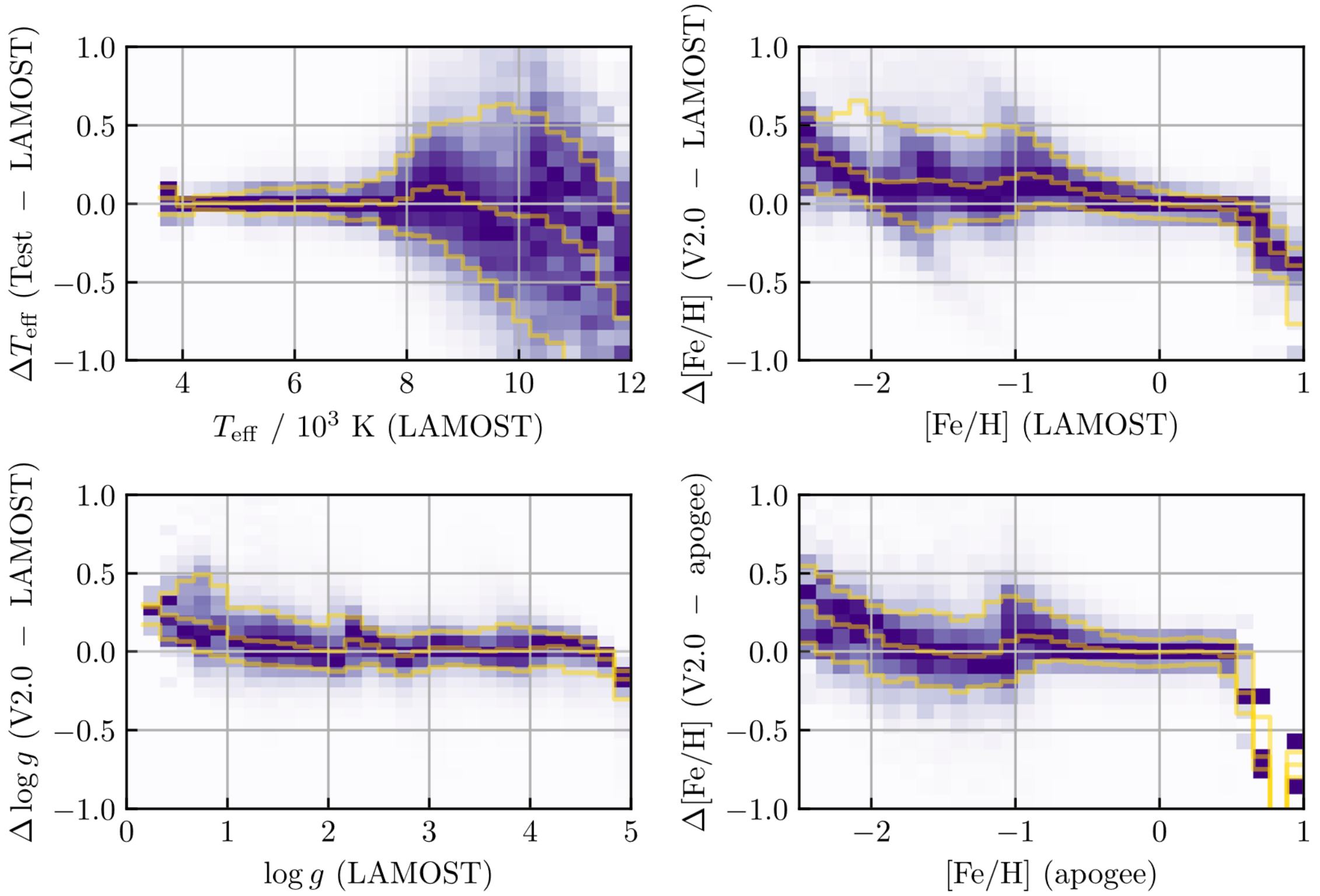

Fig. D4 Validation of our stellar parameters with relatively higher-resolution spectroscopic surveys. Comparisons with LAMOST are made in the validation set. We select stars that pass the basic cut and the corresponding confidence cut, and show their residuals of parameters as a function of LAMOST (APOGEE) parameters. The yellow lines mark the positions of 16th, 50th and 84th percentiles. The $T_{\mathrm{eff}}$ residuals are approximately flat for $4,000 < T_{\mathrm{eff}} < 10,000$ K, but the scattering increases when $T_{\mathrm{eff}} > 9,000$ K. Our $\log g$ estimates agree with LAMOST for $1 \lesssim \log g \lesssim 5$, but deviate from it when $\log g < 1$. The residual of $[\mathrm{Fe/H}]$ against LAMOST is mostly flat, but agrees better with LAMOST at $-1 < [\mathrm{Fe/H}] < 0.5$ . However, our $[\mathrm{Fe/H}]$ estimates agree with APOGEE at $-2 < [\mathrm{Fe/H}] < 0.5$, with a bump at $[\mathrm{Fe/H}] \sim= 1$, which is due to the systematics between LAMOST and APOGEE pipelines.

**Fig. S5. (D5)**

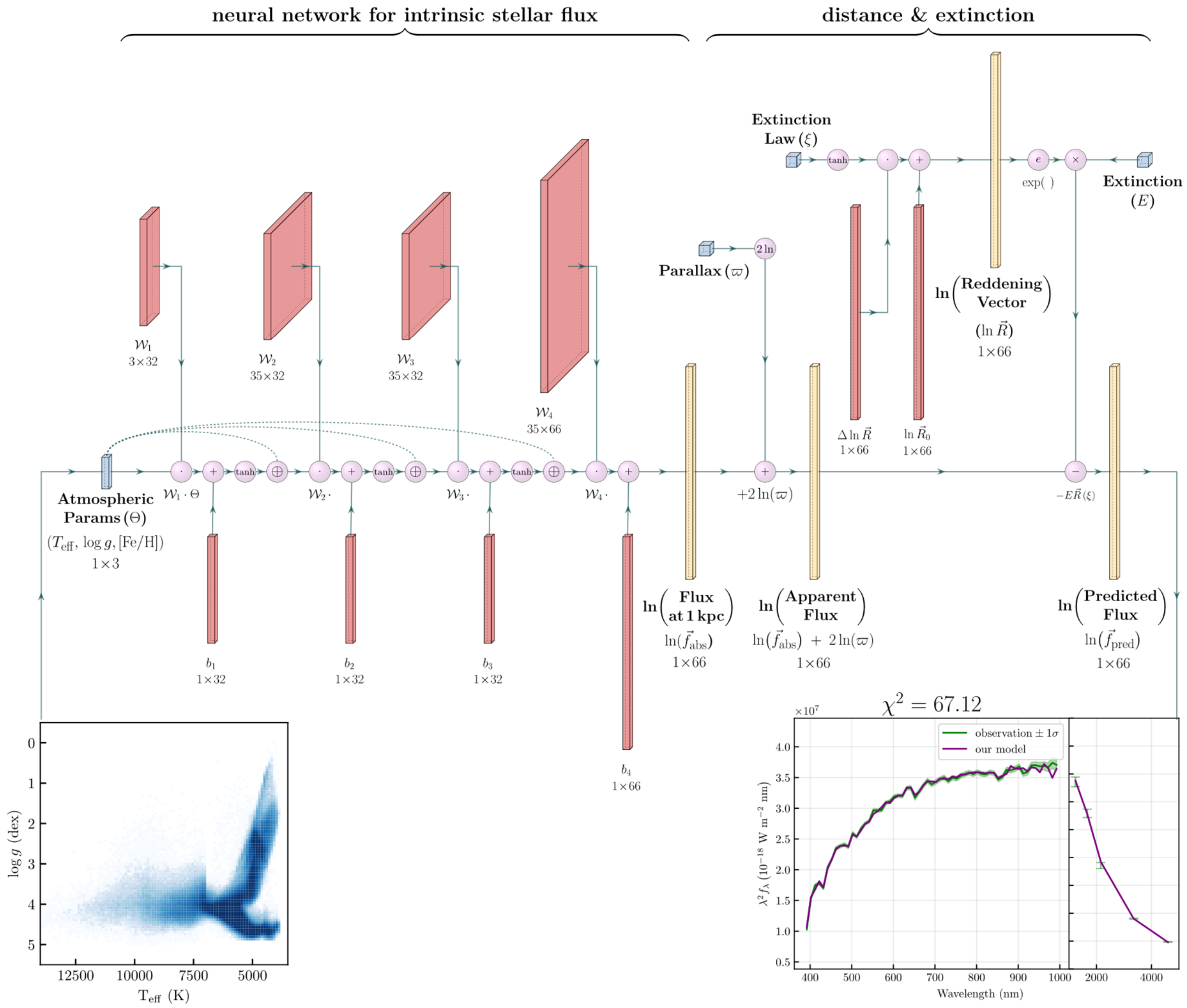

Fig. D5 The structure of the stellar flux model consists of an "intrinsic" part that predicts the dereddened flux and a physical part that accounts for distance and extinction. Red blocks represent global parameters shared by all stars, and yellow blocks represent output or intermediate calculations. The structure of the intrinsic part is similar to v1 (*11*), which utilizes a 4-layer feed-forward neural network (with weights $\mathcal{W}$ and biases $b$) to map atmospheric parameters $\theta \equiv (T_{\text{eff}},\ [\text{Fe/H}],\ \log g)$ to the natural logarithm of the de-reddened flux at 1 kpc. The only structural change to the intrinsic part of the network, relative to v1, is that we concatenate (represented by $\bigoplus$) the atmospheric parameters ($\theta$) to each hidden layer. The effect of distance is taken into account by multiplying by the square of the model parallax ($\varpi^2$). In contrast with v1, we allow the slope of the extinction curve to vary in a one-parameter family for each star by adding a 66-dimensional term ($\mathcal{W}_\xi$) to the extinction model and giving each star an additional degree of freedom ($\xi$) to account for its distinct extinction curve. Mathematically, we model extinction curves by $\exp\left[-E \exp\left(\ln \vec{R}_0 + tanh(\xi)\ \Delta ln\vec{R}\right)\right]$. In the training process, we first fix $\xi = 0$ (and thus $\mathcal{W}_\xi$ ) and train a model with universal extinction curve similar to v1. Then we set free the extinction model to learn the variation of extinction curve. Finally, we set free all parameters to optimize the model. Please refer to Table S1 for the detailed training process.

**Fig. S6. (E6)**

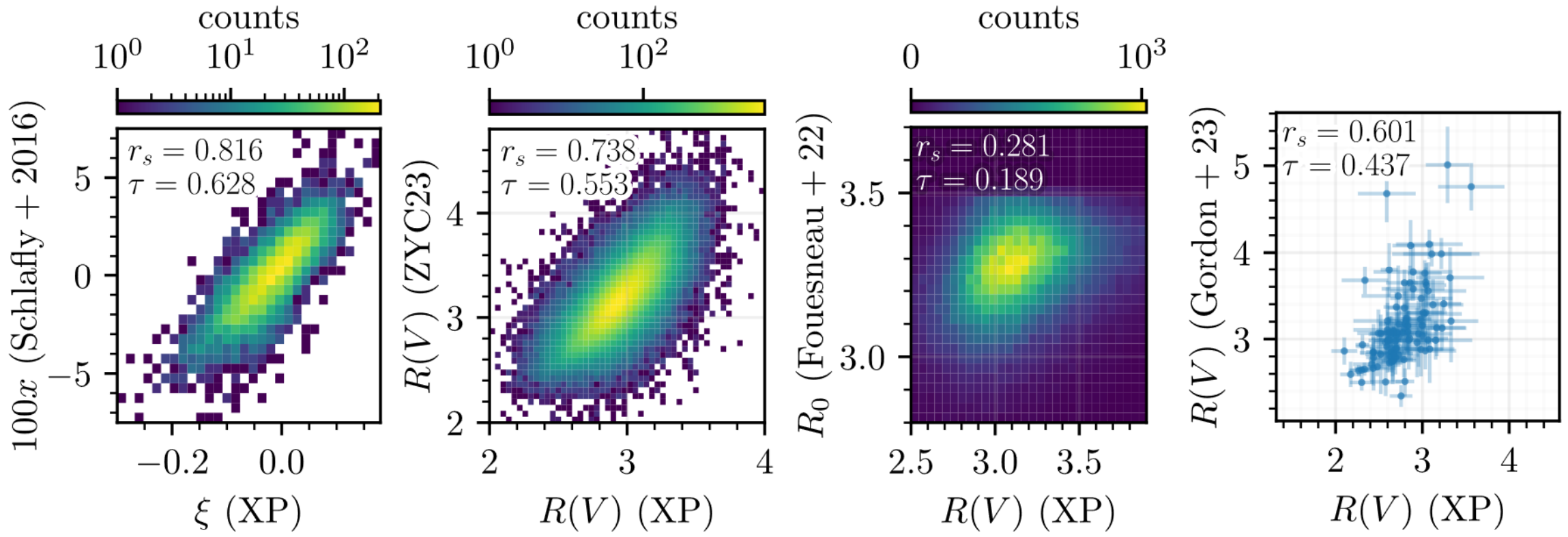

Fig. E6 Star-by-star comparisons of our extinction-curve measurements with previous catalogs. Left panel: Comparison of our inferred extinction-curve parameter, $\xi$, with the extinction-curve parameter, $x$, of Schlafly et al. (*4*, *5*, hereafter "S16"). Right three panels, from left to right: Comparison of our inferred $R(V)$ values with those of Zhang et al. (*6*, no relation; hereafter "ZYC23"), (*41*, hereafter "F22"), and (*3*, hereafter "G23"), respectively. In the right two panels, we convert our $\xi$ parameter to $R(V)$ value using our standard procedure (See Appendix B). As we do not have direct XP-based $R(V)$ estimates for the O/B stars measured by G23, we estimate the $R(V)$ of these stars using nearby XP stars (See text for details). In each panel, we additionally show the Spearman rank coefficient, $r_s$, and the Kendall $\tau$ rank coefficient, each of which indicates the level of correlation between our extinction curve measures (0 indicates no correlation, while 1 indicates perfect correlation). Different definitions of $R(V)$ can produce offsets and slope differences when comparing different catalogs, but rank coefficients allow a robust determination of whether two catalogs contain the same information. We find high correlation between our extinction curve estimates and those of S16, ZYC23 and G23, but low correlation with those of F22. Some of the scatter in our comparison with G23 derives from the fact that the O/B stars analyzed by G23 probe slightly different sightlines than the nearby stars in our catalog. In particular, O/B stars frequently probe dense regions of clouds, whereas our comparison stars probe less dense surrounding regions. Nevertheless, we find a moderately strong correlation between G23 and our estimates. We note that S16, ZYC23, G23 and our catalog make use of stellar spectra (APOGEE, LAMOST, IUE/STIS/Spitzer and Gaia XP, respectively), while F22 is based purely on broadband photometry. From our good agreement with other spectroscopic extinction curve measurements, we conclude that our estimates are robust.

**Fig. S7. (E7)**

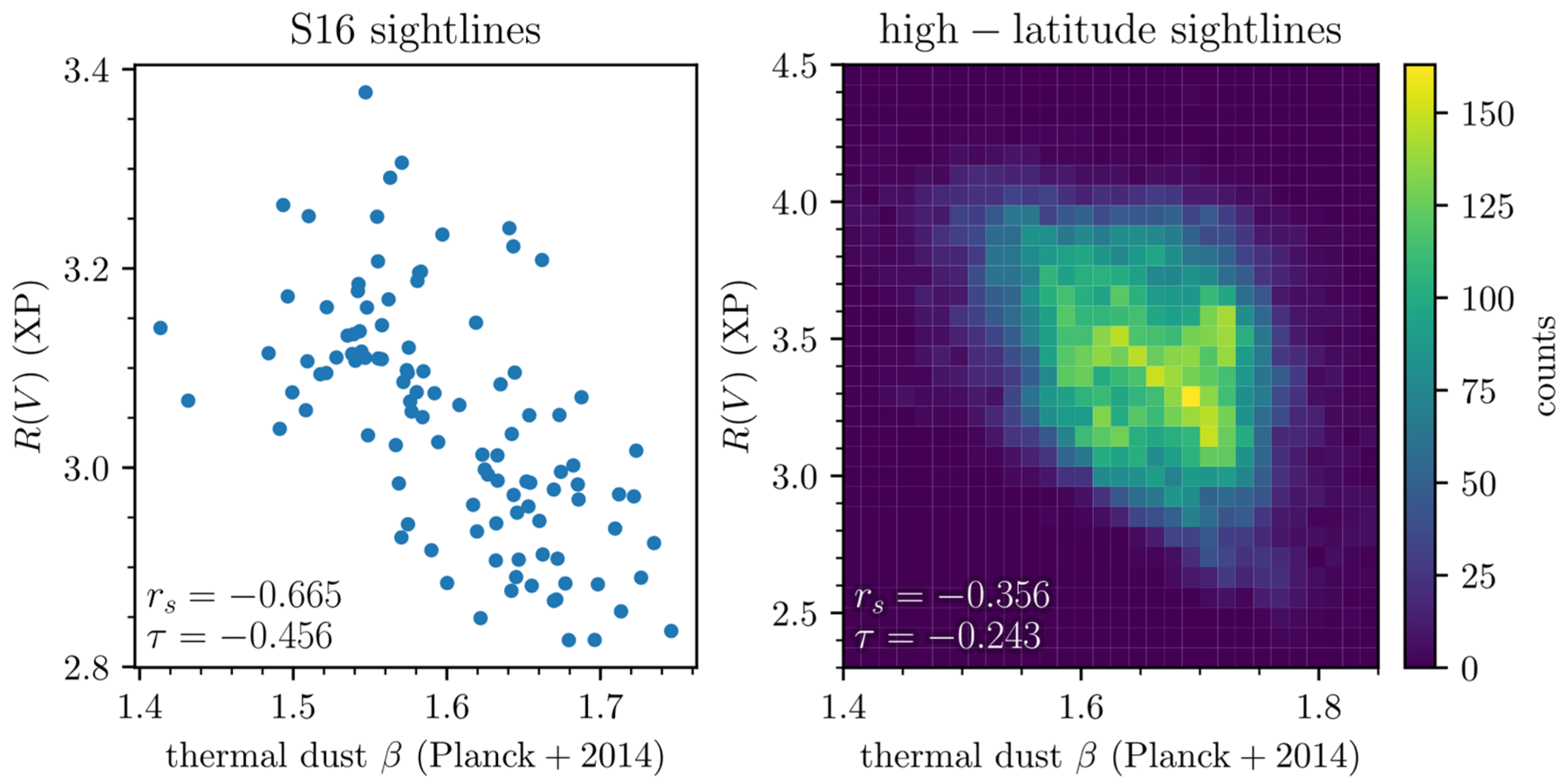

Fig. E7 Anti-correlation of $R(V)$ with the slope of the dust modified blackbody spectrum in the far-infrared (FIR), $\beta$, as measured by Planck Collaboration et al. (*36*). S16 found an anti-correlation between $\beta$ and $R(V)$ in low-Galactic-latitude sightlines. Here, we confirm S16's results for the same sightlines (left panel), and then explore possible $\beta - R(V)$ correlation at high Galactic latitudes (right panel), where we expect the stellar extinction and FIR thermal emission to more cleanly probe the same dust column. In both panels, we show both the Spearman rank ($r_s$) and Kendall $\tau$ coefficients, indicating the level of anti-correlation between $R(V)$ and $\beta$. At high Galactic latitudes, we find a smaller – yet still moderate – anti-correlation between $R(V)$ and $\beta$. This suggests that at least some of the anti-correlation between $R(V)$ and $\beta$ is intrinsic to the dust-grain population itself, rather than being a modeling artifact (e.g., of averaging multiple clouds along the line of sight).

**Fig. S8. (E8)**

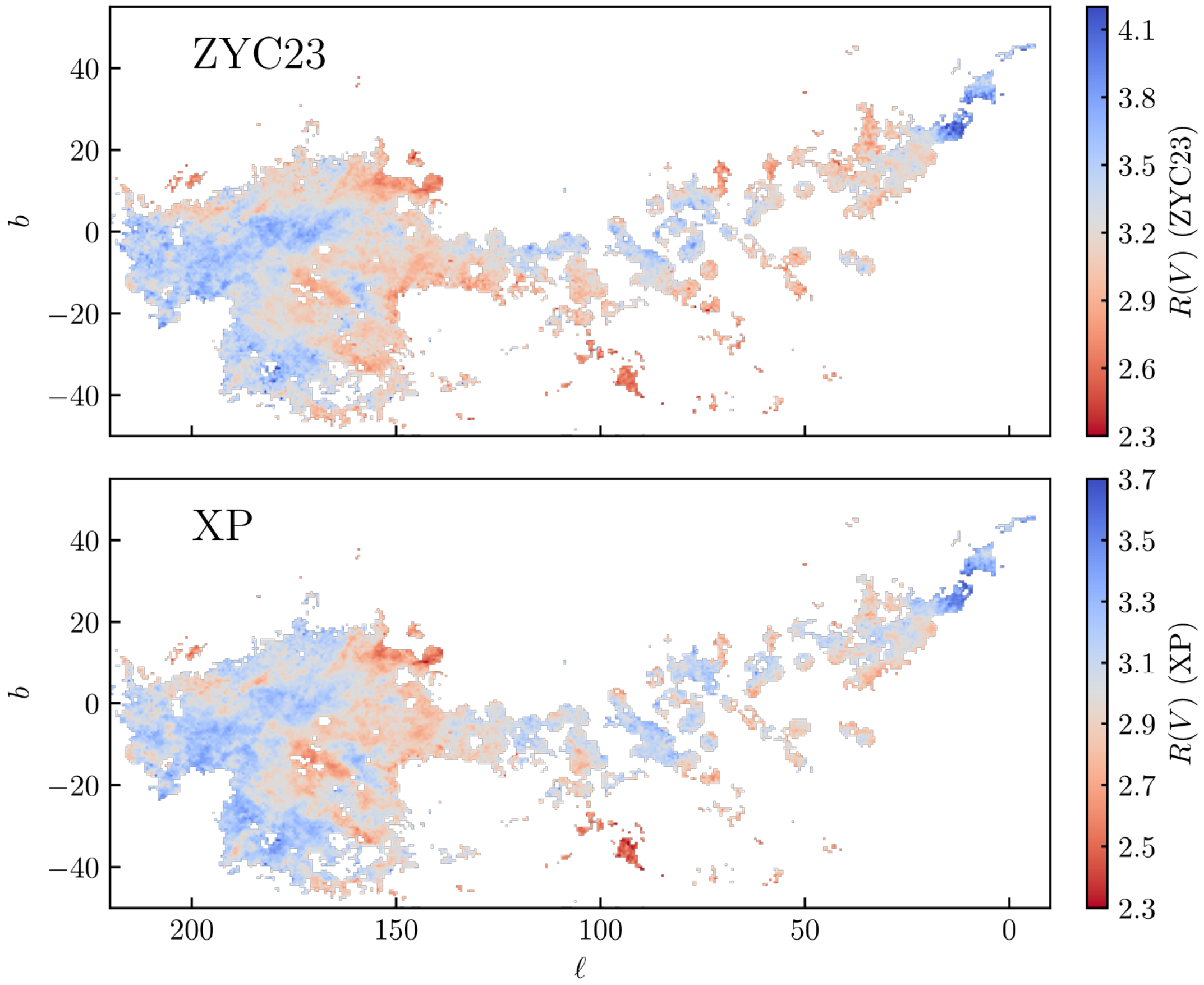

Fig. E8 2D $R(V)$ distribution compared with ZYC23 (*6*). The top panel is the 2D distribution of $R(V)$ with stars in ZYC23 that have $E > 0.1$ and $R(V)_{\text{err}} < 0.1$, similar to the Figure 6 in (*6*). Since most of the stars in ZYC23 have Gaia XP spectra, we select the common high-extinction ($E > 0.1$) stars that pass our basic cut and Teff confidence cut and show their 2D distribution in the bottom panel. Patterns of $R(V)$ in both panels are consistent, i.e., $R(V)$ is high (low) in the top panel where the same region shows high (low) $R(V)$ in the bottom panel. This is not surprising since there is a monotonically increasing (but not identical) relation between $R(V)$ in ZYC23 and this work, as shown in Fig. E6, so the colorbars are of different scales for the top and bottom panels.

**Table S1.**

| | Selection[1] | | | Weighting | | | Initial LR | |
|---|---|---|---|---|---|---|---|---|
| Phase | HQ | $E > 0.1$ | no outliers | prior | ξ hist | Variables[2] | stars | Model |
| 1 | √ | | | | | $\mathcal{M}, \vec{R}_0$ | | $10^{-4}$ |
| 2 | √ | | | | | $\theta, E, \varpi$ | $5 \times 10^{-5}$ | |
| 3 | √ | | | | | $\mathcal{M}, \vec{R}_0, \theta, E, \varpi$ | $2.5 \times 10^{-5}$ | $2.5 \times 10^{-5}$ |
| 4[3] | √ | | √ | | | $\mathcal{M}, \vec{R}_0, \theta, E, \varpi$ | $10^{-5}$ | $10^{-5}$ |
| 5 | √ | √ | | | | $\xi, E, \varpi$ | $5 \times 10^{-5}$ | |
| 6 | √ | √ | | √ | √ | $\vec{R}_0, \xi, E, \varpi$ | $5 \times 10^{-5}$ | $5 \times 10^{-5}$ |
| 7 | √ | √ | | √ | √ | $\Delta, \xi, E, \varpi$ | $5 \times 10^{-5}$ | $10^{-4}$ |
| 8 | √ | √ | | | | $\theta, \xi, E, \varpi$ | $5 \times 10^{-5}$ | |
| 9 | √ | √ | | √ | √ | all | $10^{-5}$ | $10^{-5}$ |
| 10[4] | | | | | | $E$ | $10^{-2}$ | |
| 11 | | | | | | $\theta, \xi, E, \varpi$ | $5 \times 10^{-5}$ | |
| 12 | | | √ | √ | √ | $\mathcal{M}, \vec{R}_0, \theta, \xi, E, \varpi$ | $10^{-5}$ | $10^{-5}$ |
| 13 | | | √ | √ | √ | all | $10^{-6}$ | $10^{-6}$ |

Training procedure described in Appendix D. We begin by training the model on a high-quality subset of stars, assuming the input parameters to be correct, and assuming $\xi = 0$. We then gradually allow the individual stellar parameters and the model to adjust simultaneously, until we arrive at a model with a universal extinction curve at the end of Phase 4, similar to Zhang et al. (*11*, "v1"). We then guess a simple form for the extinction curve variation ($\Delta \ln \vec{R}$), and update the stellar $\xi$ values (alongside $E$ and $\varpi$) to explain the residuals. We gradually allow more model and stellar parameters to adjust simultaneously, until we adjust all parameters in phase 9, using only high-extinction, high-quality stars. Then, we bring lower-quality stars into the training, first updating their individual parameters, and then updating both the model and stars simultaneously. In the final phase, we simultaneously train all model and stellar parameters, using stars that pass our self-cleaning outlier rejection. The resulting model has a variable extinction curve. We provide code to replicate our training procedure at https://zenodo.org/doi/10.5281/zenodo.10719756

[1]The combined selection is the intersection of all checked selections. For details on each selection, see Appendix D.

[2]Variable shorthand: M stands for the parameters in the intrinsic stellar model ($\{\mathcal{W}, b\}$); $\vec{R}_0$ refers to the mean extinction curve; Δ is shorthand for $\Delta \ln \vec{R}$, which describes the variation in the extinction curve; $\theta$, $\xi$, $E$, and $\varpi$ are the individual stellar atmospheric parameters, extinction law, extinction, and parallax, respectively.

[3]The result of phase 4 is similar to v1 of our model.

[4]In this phase, only stars with highly uncertain extinction priors are selected: $\sigma_E > 0.1$ .